%% file: perturbations230314.tex
\documentclass[11pt]{article}
\pdfoutput=1  
\usepackage{graphicx,color}
\usepackage{appendix}
\usepackage{latexsym,amsmath,amssymb,graphicx,booktabs}
\usepackage{epsfig,latexsym,cite}
\usepackage{hyperref}
\numberwithin{equation}{section}

\definecolor{MyBlue}{rgb}{0.15,0.15,0.70}

\hypersetup{
colorlinks=true,
citecolor=MyBlue,
linkcolor=MyBlue,
urlcolor=MyBlue
}

\input epsf
\usepackage{amssymb}
\usepackage{amsmath}
\usepackage{amsfonts}
\usepackage{upgreek}
\usepackage{latexsym}

\include{mydefs}

\begin{document}

\begin{titlepage}

\vspace*{2cm}

\centerline{\Large \bf Cosmological perturbations and structure formation\\}

\vspace{5mm}

\centerline{\Large \bf in nonlocal infrared modifications of general relativity}

\vskip 0.4cm
\vskip 0.7cm
\centerline{\large Yves Dirian$^1$, Stefano Foffa$^1$, Nima Khosravi$^2$, Martin Kunz$^{1,3}$ and  Michele Maggiore$^1$}
\vskip 0.3cm
\centerline{\em $^1$D\'epartement de Physique Th\'eorique and Center for Astroparticle Physics,}  
\centerline{\em Universit\'e de Gen\`eve, 24 quai Ansermet, CH--1211 Gen\`eve 4, Switzerland}
\vspace{3mm}
\centerline{\em $^2$School of Astronomy, Institute for Research in Fundamental Sciences (IPM),}
\centerline{\em  P. O. Box 19395-5531, Tehran, Iran }  
\vspace{3mm}
\centerline{\em $^3$African Institute for Mathematical Sciences, 6 Melrose Road, } 
\centerline{\em Muizenberg, 7945, South Africa}

\vskip 1.9cm

\begin{abstract}
We study the cosmological consequences of a recently proposed nonlocal modification of general relativity, obtained by adding a term $m^2R\,\Box^{-2}R$ to the Einstein-Hilbert action. The model has the same number of parameters as $\Lambda$CDM, with $m$ replacing $\Omega_{\Lambda}$, and is very predictive. 
At the background level, after fixing $m$ so as to reproduce the observed value of $\Omega_M$, we get a pure prediction for the equation of state of dark energy as a function of redshift, $w_{\rm DE}(z)$, with 
$w_{\rm DE}(0)$ in the range $[-1.165,-1.135]$ as $\Omega_M$ varies over  the broad range $\Omega_M\in [0.20,0.36]$. We find that the cosmological perturbations are well-behaved,  and the model fully fixes  the dark energy perturbations as a function of redshift $z$ and wavenumber $k$.
The  nonlocal model provides a good fit to supernova data and predicts deviations from General Relativity in  structure formation and in weak lensing at the level of 3-4\%, therefore consistent with existing data but readily detectable by future surveys. For the logarithmic growth factor we obtain $\gamma\simeq 0.53$, to be compared with 
$\gamma\simeq 0.55$ in $\Lambda$CDM. For the Newtonian potential  on subhorizon scales    our results are well fitted by
$\Psi(a;k)=[1+\mu_s a^s]\Psi_{\rm GR}(a;k)$
with a scale-independent $\mu_s\simeq 0.09$ and $s\simeq 2$, while the anisotropic stress is negligibly small.

\end{abstract}


\end{titlepage}

\newpage

\section{Introduction}

The problem of understanding the origin of dark energy (DE) has stimulated in recent years a very active search for modifications of General Relativity (GR). The challenge is to construct a theoretically consistent theory that modifies GR in the far infrared, i.e.  at cosmological scales, while retaining its successes at the scale of the solar system and of terrestrial laboratories. The first example of an infrared  modification of GR was provided by the DGP model \cite{Dvali:2000hr}, which indeed has a  self-accelerated solution \cite{Deffayet:2000uy,Deffayet:2001pu}. This solution is however   plagued by a 
ghost instability~\cite{Luty:2003vm,Nicolis:2004qq,Gorbunov:2005zk,Charmousis:2006pn,Izumi:2006ca} and is therefore not viable.  Significant advances have then been done toward the construction of a consistent theory of massive gravity with the dRGT theory
\cite{deRham:2010ik,deRham:2010kj} (see also \cite{deRham:2010gu,deRham:2011rn,deRham:2011qq,Hassan:2011hr,Hassan:2011vm,Hassan:2011tf,Hassan:2011ea,Hassan:2012qv,Comelli:2012vz,Comelli:2013txa,Jaccard:2012ut}), although at present a number of open conceptual issues still persist, and it is also unclear whether acceptable cosmological solutions emerge
(see \cite{Hinterbichler:2011tt,deRham:2014zqa} for  reviews).

In a recent series of papers \cite{Jaccard:2013gla,Maggiore:2013mea,Foffa:2013sma,Foffa:2013vma,Kehagias:2014sda,Maggiore:2014sia}  an alternative approach has been proposed  in which a mass parameter enters the theory as the coefficient of a  nonlocal term.  
Different implementations of the idea have been explored. The one which is probably closest in spirit to the original degravitation idea~\cite{ArkaniHamed:2002fu,Dvali:2006su} consists in writing a modified Einstein equation of the form 
\be\label{modelGmnT}
\Gmn -m^2\(\iBox_g\Gmn\)^{\rm T}=8\pi G\,\Tmn\, ,
\ee
where the superscript T denotes the operation of taking the transverse part (which is itself a nonlocal operation), $\Box_g$ is the d'Alembertian computed with the curved-space metric $\gmn$, and its inverse
$\iBox_g$ is defined using the retarded Green function. The extraction of the transverse part ensures that energy-momentum conservation is still automatically satisfied (see also
\cite{Porrati:2002cp}), while the use of a retarded Green's function ensures causality.
It was then realized in 
\cite{Maggiore:2013mea,Foffa:2013vma,Modesto:2013jea} that such tensor nonlocalities generate instabilities in the cosmological evolution (see also \cite{Ferreira:2013tqn} for similar conclusions in a different nonlocal model). The attention then shifted to theories where the nonlocal operator $\iBox$ is applied to the Ricci scalar. Basically, two possibilities come to mind. One possibility, which was  proposed in
\cite{Maggiore:2013mea}, is to add a term  $m^2(\gmn\iBox_g R)^ {\rm T} $  to the Einstein equations,
writing
\be\label{modelRT}
\Gmn -(1/3)m^2\(\gmn\iBox_g R\)^{\rm T}=8\pi G\,\Tmn\, ,
\ee
(where the factor $1/3$ is a convenient normalization of the parameter $m^2$ in $d=3$ spatial dimensions). We will refer to this as the ``$\gmn\iBox R$ model".
It was found in \cite{Maggiore:2013mea} that this model   
generates a dynamical dark energy. Its value today can be matched to the observed value 
$\ode\simeq 0.68$ by tuning the mass $m$ (which is obtained setting $m\simeq 0.67 H_0$). The fact that a DE is dynamically generated and that the observed value can be reproduced is already quite significant. Furthermore, having fixed $m$, we have fixed the only free parameter of the theory and we then obtain  a pure prediction for the EOS parameter of dark energy. For this model, writing in the recent epoch
$w_{\rm DE}(a)=w_0+(1-a) w_a$,  one finds 
$w_0\simeq-1.04$ and $w_a\simeq -0.02$ \cite{Maggiore:2013mea}, which is consistent with the {\it Planck} data, and on the phantom side.
In an  interesting recent paper, Nesseris and Tsujikawa \cite{Nesseris:2014mea} have studied the cosmological perturbations of this model and have compared them to CMB, BAO, SNIa and growth rate data. They find that, if one uses a prior on $H_0$ derived from local measurements of the Hubble parameter \cite{Riess:2011yx}, $h_0\,\gsim\, 0.70$,  the data strongly support this nonlocal model over $\Lambda$CDM, while using a lower prior,
$0.67\lsim\, h_0\,\lsim\, 0.70$, as suggested by the {\it Planck} data \cite{Ade:2013zuv}, the two models are statistically comparable.
It should be observed that the nonlocal gravity model has the same number of free parameters as 
$\Lambda$CDM, with the mass $m$ replacing $\ola$.

A second possibility, recently put forward in \cite{Maggiore:2014sia}, is to add a term involving $R\,\Box_g^{-2} R$ directly to the action, writing 
\be\label{S1}
S_{\rm NL}=\frac{1}{16\pi G}\int d^{4}x \sqrt{-g}\, 
\[R-\frac{1}{6} m^2R\frac{1}{\Box_g^2} R\]\, .
\ee
Observe that, upon integration by parts, we can equivalently write $R\,\Box^{-2}R=(\iBox R)^2$.
We will refer to it as the ``$R\,\Box^{-2}R$ model".
As discussed in  \cite{Maggiore:2014sia}, when linearizing the equations of motion derived from the action (\ref{S1}) around flat space, one finds the same equations of motion as those obtained by linearizing \eq{modelRT}. However, at the full non-linear level, the two theories are different. In the $R\,\Box^{-2}R$ model there is again a dynamically generated dark energy, which can be made to agree with the presently observed value by choosing $m\simeq 0.28H_0$. The prediction for the DE equation of state is then
$w_0\simeq -1.14$, $w_a=0.08$ (with a mild dependence on the value of $\Omega_M$ today, that will be discussed in more detail below). These values are compatible with existing limits, but will be easily distinguished from the predictions of $\Lambda$CDM with forthcoming data. In particular, in the next few years the DES survey should measure $w_0$ to an accuracy of about
$\Delta w_0\simeq 0.03-0.04$ and later {\sc Euclid} should measure it  to an accuracy
$\Delta w_0\simeq 0.01$ \cite{Amendola:2012ys}. The above models are therefore highly testable.
In this paper we focus in particular in the $R\,\Box^{-2}R$ model. Since its predictions,  at the level of background evolution (and, as we will see in this paper, also at the level of perturbations), differ from $\Lambda$CDM more than the predictions of the
$\gmn\iBox R$ model, it is presumably the first of the two that will be ruled out (or possibly confirmed) by future data.

At the conceptual level, one might be worried by the presence of nonlocal terms in the equations of motion. However, 
it is important to observe that nonlocal classical equations, constructed  with a retarded Green function, appear in a number of different situations. As discussed in detail in 
\cite{Maggiore:2013mea,Foffa:2013sma} (and as recognized in similar contexts also in \cite{Tsamis:1997rk,Deser:2007jk,Barvinsky:2011rk,Deser:2013uya}), such nonlocal equations
should not be thought of as the classical equations of motion of a fundamental nonlocal quantum field theory. Rather, they can emerge, already in a purely classical context, from some form of smoothing or iterative procedure in an underlying local fundamental theory.  Probably the simplest example is provided by the formalism for gravitational-wave production in GR beyond lowest order.  In linearized theory the gravitational wave (GW) amplitude $\hmn$ is determined by  
$\Box\bhmn=-16\pi G\Tmn$, where $\bhmn=\hmn-(1/2)h\emn$.
In such a classical radiation problem, this equation is solved with the retarded Green's function,  
$\bhmn=-16\pi G\iBox_{\rm ret}\Tmn$. When the non-linearities of GR are included,
the GWs generated at some perturbative order become themselves sources for the GW generation at the next order. In the far-wave zone, this iteration gives rise to effective nonlocal equations involving $\iBox_{\rm ret}$, and  is at the basis of both the  Blanchet-Damour and the Will-Wiseman-Pati  formalisms 
(see e.g. \cite{Blanchet:2006zz} or chapter~5 of \cite{Maggiore:1900zz} for reviews).
A nonlocal action can be seen as a compact way of summarizing such effective nonlocal equations of motion.\footnote{However, the use of a nonlocal action implies a (rather revealing) 
subtlety~\cite{Deser:2007jk,Barvinsky:2011rk,Jaccard:2013gla,Foffa:2013sma}. The variation of  a nonlocal action involving $\iBox$, where $\iBox$ is defined with some Green's function $G(x;x')$, produces equations of motion where appears $\iBox$ constructed with the symmetrized  Green's function $(1/2)[G(x;x')+G(x';x)]$. It is therefore impossible to obtain in this way a retarded Green's function in the equations of motion. We can still take the formal variation of the action and at the end replace by hand all factors $\iBox$ by
$\iBox_{\rm ret}$ in the equation of motion. In this way, the nonlocal action is seen just 
as a convenient ``device" that allows us to compactly summarize   the equations of motion.
However, any connection to a corresponding nonlocal quantum field theory is then lost.
Indeed, also the action (\ref{S1}) should be understood in this sense. In other words, the nonlocal classical theory that we consider  is {\em defined} by the equations of motion derived from a formal variation of \eq{S1}, in which $\iBox\ra\iBox_{\rm ret}$. The use of an action is however convenient because it ensures automatically the covariance of the equations of motions.} Another example of this type is the effective action describing the interaction between two compact bodies in GR, which at fourth post-Newtonian order develops a   term nonlocal in time \cite{Damour:2014jta}. Such a term reflects the existence of the so-called ``tail terms", i.e. nonlocal terms that represent radiation emitted earlier and that come back to the particle after performing multiple scattering on the background curvature. Such terms therefore depend on the whole past history (see also \cite{Poisson:2011nh}). One more recent example of this type is the effective field theory of  cosmological perturbations, which is an effective classical theory
for the long-wavelength modes obtained   by integrating 
out the short-wavelength modes~\cite{Baumann:2010tm}
and  again has terms that are nonlocal in time, expressed through a retarded Green function~\cite{Carrasco:2012cv,Carroll:2013oxa}. The above examples are purely classical.
Nonlocal effective classical equations can also appear by performing a quantum averaging. Nonlocal field equations govern the effective dynamics of the vacuum expectation values of quantum fields.  In particular, the in-in matrix elements of operators satisfy nonlocal but causal 
equations, involving only  retarded propagators \cite{Jordan:1986ug,Calzetta:1986ey}.
The bottom-line of this discussion is that nonlocality often appears in physics, but is always derived from some averaging process in a fundamental local theory. Issues of quantum consistency, such as the possible existence of ghosts in the spectrum of the quantum theory,
cannot be addressed in the effective nonlocal classical theory, but can only be studied in the underlying fundamental quantum theory. 

The approach that we are proposing, based on the addition of nonlocal terms to the Einstein equations, therefore has two natural directions of development:
(1) to understand whether such nonlocal effective classical equations can be embedded in a consistent quantum  theory, and (2): to understand whether such models have interesting and viable cosmological consequences. At least in a first approximation these two problems 
are decoupled.\footnote{Of course, one must keep in mind the possibility that the necessity of embedding the classical equations in a consistent quantum theory will require a different nonlocal structure.  In any case, the study of the cosmological consequences of models such as (\ref{S1}) will provide a first step for further refinements.
Also, in principle the fundamental quantum theory could have degrees of freedom that modify, e.g., the spectrum of quantum fluctuations at inflation that seed the subsequent cosmological evolution. This would affect the prediction of a fundamental inflationary model 
for the spectral index $n_s$ and the amplitude of the gravitational potential $\d_H$, that appear in the initial conditions, see \eqst{initcond1}{initcond4}. In any case, in a full analysis obtained evolving the perturbations with a Boltzmann code and comparing with the data,  $n_s$ and  $\d_H$ will be taken as free parameters to be fitted, just as one does in $\Lambda$CDM.}
In this paper we study the cosmological perturbations of these nonlocal cosmological models, focusing in particular on the 
$R\,\Box^{-2}R$ model as our reference model. The first issue that we wish to understand is whether the perturbations are well-behaved. This is already a non-trivial point. Indeed,  infrared extensions of GR such as DGP have been ruled out by the lack of well-behaved perturbations over the cosmologically interesting solutions, and similar  problems can appear in massive gravity theories such as dRGT
\cite{Gumrukcuoglu:2011ew,Gumrukcuoglu:2011zh,Koyama:2011wx}. We will see that, in our nonlocal model, cosmological perturbations are indeed well behaved. 
This opens the way for a more detailed comparison with  CMB, BAO, SNIa and structure formation, and we will see that the model performs quite well when compared   to  observations. 

Finally, we observe that our model differs from the nonlocal model  proposed by Deser and Woodard
\cite{Deser:2007jk,Deser:2013uya,Woodard:2014iga}
and studied in many subsequent papers (see e.g. \cite{Nojiri:2007uq,Jhingan:2008ym,Koivisto:2008xfa,Koivisto:2008dh,Capozziello:2008gu,Elizalde:2011su,Zhang:2011uv,Elizalde:2012ja,Bamba:2012ky,Park:2012cp,Dodelson:2013sma}, and also \cite{Barvinsky:2003kg,Barvinsky:2011hd,Barvinsky:2011rk} for a related approach). 
The Deser-Woodard model  does not involve a mass scale $m$, and is instead constructed adding to the Einstein-Hilbert action a term of the form $Rf(\iBox R)$. The function $f(\iBox R)$ is then tuned so that, at the level of background evolution, this model reproduces $\Lambda$CDM, which turns out to require that 
$f(X)=a_1[\tanh (a_2Y+a_3Y^2+a_4Y^3)-1]$ with $Y=X+a_5$, and $a_1,\ldots a_5$ suitably chosen coefficients.  The action of this model is therefore significantly more involved, compared to the nonlocal  action (\ref{S1}) where, in terms of $X=\iBox R$, the nonlocal term is simply $m^2 X^2$, and is also not predictive as far as the background evolution is concerned. More importantly, after fixing 
$f(\iBox R)$ so to reproduce the background evolution of $\Lambda$CDM, one can study its
cosmological perturbations, and  it has been found in 
\cite{Dodelson:2013sma} that the Deser-Woodard  model is ruled out  by comparison with structure formation (with the model being disfavored, with respect to GR, at the 7.8~$\sigma$ level from redshift space distortion, and at the 5.9~$\s$ level from weak lensing). This shows the power of 
structure formation data for testing nonlocal modifications of GR, and it is therefore natural to ask how our nonlocal models perform in this respect. We will see that
the model (\ref{S1}) (and also the model (\ref{modelRT}), as recently shown in \cite{Nesseris:2014mea}) passes these tests with flying colors, giving predictions for structure formation that are sufficiently close to 
$\Lambda$CDM to be consistent with existing data, yet sufficiently different to be distinguishable by near-future surveys. We also observe that a phantom equation of state has also been obtained recently in
\cite{Konnig:2014dna} for a bimetric gravity model. In this case, $w_{\rm DE}(0)\simeq -1.22$ and again structure formation is consistent with existing data, with a growth index $\gamma\simeq 0.47$.

The organization of the paper is as follows. In sects.~\ref{sect:model} and \ref{sect:back} we recall the properties of the model and we study its background evolution, expanding on the results already presented in \cite{Maggiore:2014sia}. Since supernova (SN) data are mostly sensitive to the background evolution, the results found in these sections  already allow us to test the nonlocal model against SN data, and we find that it performs as well as $\Lambda$CDM, although the fit to SN data suggests a higher value of $\oma$, compared to $\Lambda$CDM.
The equations governing the cosmological perturbations for the $R\,\Box^{-2}R$ model are presented in sect.~\ref{sect:pert}.  In sect.~\ref{sect:analytic} we derive analytic results in the sub-horizon limit, and we show that the predictions of this model are well compatible with  the data on structure formation. We confirm this discussion in sect.~\ref{sect:numeric} by numerically integrating the perturbation equations, and we also show the full evolution on all scales.
Sect.~\ref{sect:concl} contains our conclusions. In App.~\ref{app:gmnR} we collect similar results for the perturbations of the 
$\gmn\iBox R$ model, that partially overlap with those recently presented in
\cite{Nesseris:2014mea}.

\section{The model}\label{sect:model}

We consider the model defined by the action (\ref{S1}), in $d=3$ spatial dimensions. 
We introduce the auxiliary fields $U$ and $S$ from
\bees
U &=& -\iBox_g R\, ,\label{Udef}\\
S  &=& - \iBox_g U\, .\label{Sdef}
\ees
Then the equations of motion are \cite{Maggiore:2014sia}
\bees
G^\mu_\nu - \frac{1}{6}m^2  K^\mu_\nu &=& 8 \pi G \,T^\mu_\nu\, ,\label{eqmRiBox2R} \\
\Box_gU&=&-R\, ,\label{BoxU}\\
\Box_gS&=&-U\, ,\label{BoxS}
\ees
where
\begin{align}
K^\mu_\nu \equiv 2 S G^\mu_\nu - 2 \nabla^\mu \partial_\nu S + 2 \delta^\mu_\nu \Box_g S + \delta^\mu_\nu \partial_\rho S \partial^\rho U - \frac{1}{2} \delta^\mu_\nu U^2 - \big( \partial^\mu S \partial_\nu U + \partial_\nu S \partial^\mu U \big).
\end{align}
We have rewritten \eq{eqmRiBox2R} in $(1,1)$-tensorial form, which will be convenient for  its perturbative treatment. 
Observe that, since the left-hand side of \eq{eqmRiBox2R} derives from the variation of a covariant action, it is transverse by construction. Indeed, using the equations of motion of the auxiliary fields it is straightforward to check explicitly that $\n_{\mu}K^{\mu}_{\nu}=0$, so the energy-momentum  $T^{\mu}_{\nu}$ is automatically conserved,
\be\label{EMcons}
\nabla_\mu T^{\mu}_{\nu} = 0\, . 
\ee
The introduction of the auxiliary fields $U$ and $S$ is technically convenient since it allows us to rewrite the original nonlocal model  (\ref{S1}) as a set of coupled differential equations. However, as discussed in detail in 
\cite{Maggiore:2013mea,Foffa:2013vma, Foffa:2013sma} for this model (and  in 
\cite{Koshelev:2008ie,Koivisto:2009jn,Barvinsky:2011rk,Deser:2013uya} for similar nonlocal models), these formulations are not equivalent, and the space of solutions of the local formulation is larger than that of the original nonlocal model.
This originates from the fact that the kernel of the  $\Box_g$ operator is  non-trivial, since the equations
$\Box_g U=0$ and $\Box_g S=0$ do not have only $U=0$ and, respectively, $S=0$ as solutions. To understand and illustrate the consequences of this fact consider for example the inversion of the $\Box_g$ operator in an unperturbed flat FRW metric, $ds^2=-dt^2+a^2(t)d\vx^2$.  The d'Alembertian operator on a scalar function $f$ is given by $\Box_g f=-a^{-d}\pa_0(a^d\pa_0f)$. A possible inversion is then given 
by \cite{Deser:2007jk}
\be\label{iBoxDW}
(\iBox_g  R)(t)=-\int_{t_*}^{t} dt'\, \frac{1}{a^d(t')}
\int_{t_*}^{t'}dt''\, a^d(t'') R(t'')\, ,
\ee
where $t_*$ is some initial value of time. With this definition, $U\equiv -\iBox_g R$ is such that $U(t_*)=0$ and $U'(t_*)=0$, so the initial conditions on $U$ are fixed once we specify what we mean by $\iBox_g R$. In other words, the space of solutions of the local system of equations (\ref{eqmRiBox2R})-(\ref{BoxS}), which corresponds to arbitrary initial conditions for $S$ and $U$, is much larger than the space of solutions of the original nonlocal model. To recover the solutions of the original nonlocal model, we must impose suitable boundary conditions on $U$ and $S$.

More generally, we could define $\iBox_g$ such that
\be
U(t)\equiv  -\iBox_g R 
\equiv U_{\rm hom}(t)+\int_{t_*}^{t} dt'\, \frac{1}{a^d(t')}
\int_{t_*}^{t'}dt''\, a^d(t'') R(t'')\, ,
\ee
where $U_{\rm hom}(t)$ is a given solution of $\Box U=0$. In any case, the point is that each definition of the $\iBox_g$ operator, i.e. each definition of the original nonlocal theory, corresponds to one and only one choice of the homogeneous solution and therefore of the initial conditions for $U$. In this paper we will study the nonlocal model that corresponds to taking initial conditions $U=0$ and $S=0$ deep in the radiation dominated (RD) era, which corresponds to the ``minimal" model studied in 
\cite{Maggiore:2013mea,Foffa:2013vma}. Observe that, deep into the RD era, the Ricci scalar $R$ vanishes, so the definition (\ref{iBoxDW}) becomes independent of the time $t_*$. Observe also that the retarded prescription in the inversion of $\Box_g$ in \eqs{Udef}{Sdef} is automatically taken into account in the local formulation, by assigning the initial condition at a reference time $t_*$ and integrating the differential equations forward in time.

\section{Background evolution}\label{sect:back}

\subsection{Evolution equations}

\begin{figure}[t]
\centering
\includegraphics[width=0.45\columnwidth]{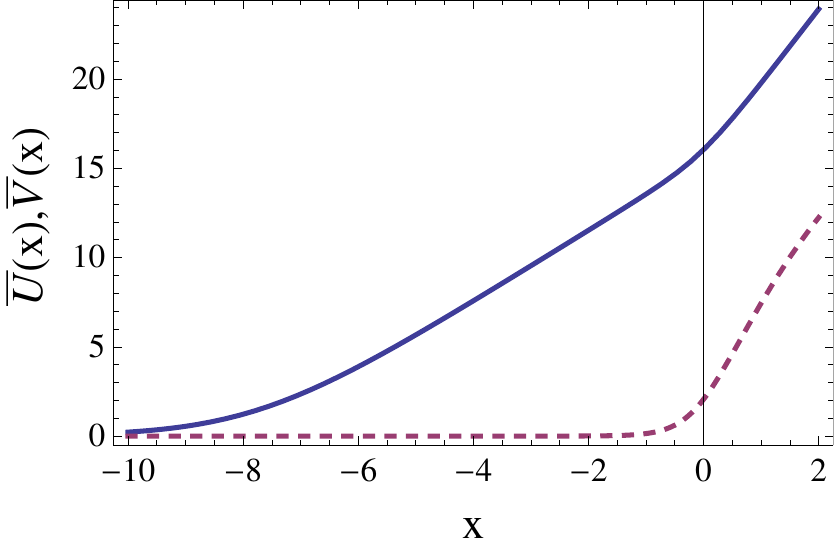}
\caption{\label{fig:UV} The functions $\bar{U}(x)$ (blue solid line) and $\bar{V}(x)$ (red dashed line), choosing $\gamma\simeq 0.0089247$.}
\end{figure}

\begin{figure}[t]
\centering
\includegraphics[width=0.45\columnwidth]{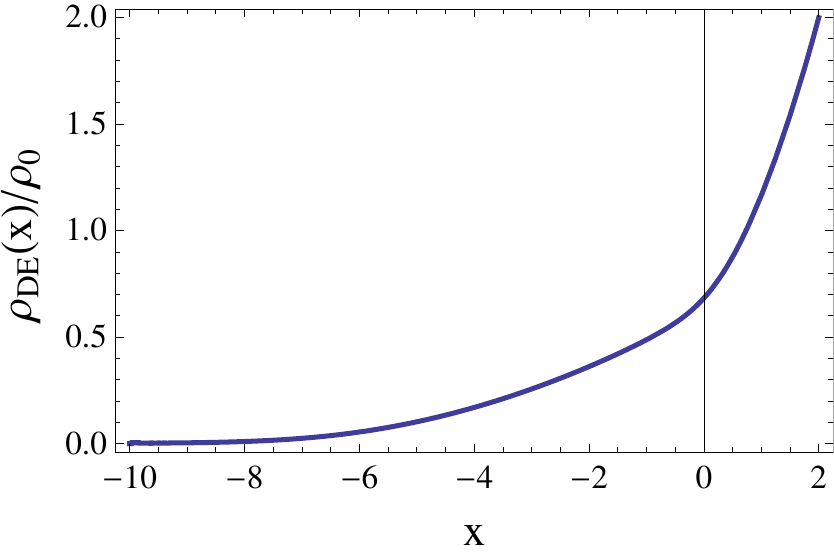}
\includegraphics[width=0.45\columnwidth]{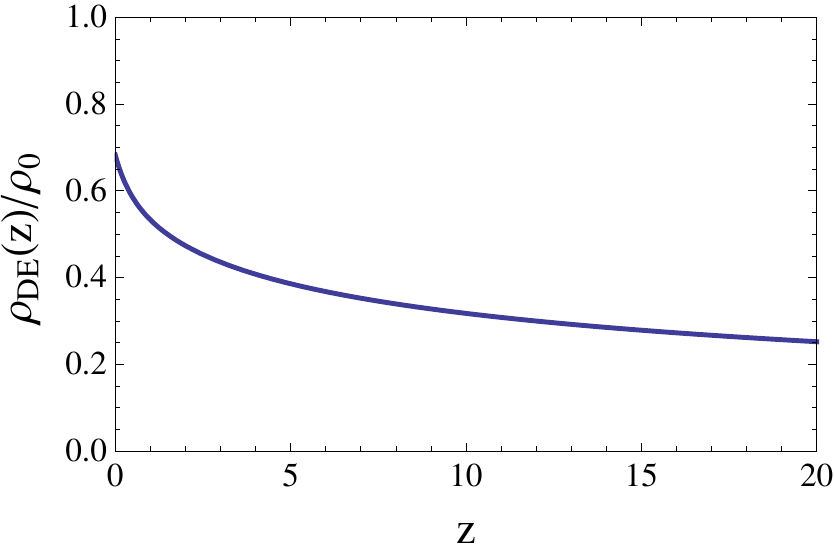}
\caption{\label{fig:rhoDE} Left panel: the function $\rde(x)/\rho_0$  agains $x=\ln a$.
Right panel:   the function $\rde(z)/\rho_0$, plotted now as a function of the redshift $z=e^{-x}-1$.
}
\end{figure}

We consider a flat FRW metric  
\be
ds^2=-dt^2+a^2(t)d\vx^2\, ,
\ee 
in  $d=3$. We use an overbar to denote the background values of $U$ and $S$, and introduce $\bar{W}(t)=H^2(t)\bar{S}(t)$ and $h(t)=H(t)/H_0$, where  $H(t)=\dot{a}/a$ and $H_0$ is the present value of the Hubble parameter. We use $x=\ln a$ to parametrize the temporal evolution, and we denote $df/dx\equiv f'$. From the $(00)$ component of
\eq{eqmRiBox2R}, together with \eqs{BoxU}{BoxS}, we get \cite{Maggiore:2014sia}
\bees
&&h^2(x)=\Omega_M e^{-3x}+\Omega_R e^{-4x}+\g \bar{Y}\, ,\label{syh}\\
&&\bar{U}''+(3+\zeta) \bar{U}'=6(2+\zeta)\, ,\label{syU}\\
&&\bar{W}''+3(1-\zeta) \bar{W}'-2(\zeta'+3\zeta-\zeta^2)\bar{W}= \bar{U}\, ,\label{syW}
\ees
where $\oma, \ora$ are the present values of $\rho_M/\rho_{\rm tot}$ and $\rho_R/\rho_{\rm tot}$, respectively, 
$\gamma= m^2/(9H_0^2)$, $\zeta=h'/h$ and
\be
\bar{Y}\equiv \frac{1}{2}\bar{W}'(6-\bar{U}') +\bar{W} (3-6\zeta+\zeta \bar{U}')+\frac{1}{4}\bar{U}^2\, .
\ee
In this form, one sees that  there is an effective dark energy density $\rde=\rho_0\gamma \bar{Y}$ where, as usual, 
$\rho_0=3H_0^2/(8\pi G)$. Actually, to perform the numerical integration of these equations and also to study the perturbations, it can be more convenient to use a variable $\bar{V}(t)=H_0^2\bar{S}(t)$ instead of $\bar{W}(t)=H^2(t)\bar{S}(t)$. Then 
\eqst{syh}{syW} are replaced by
\bees
&&h^2(x)=\frac{\Omega_M e^{-3x}+\Omega_R e^{-4x}+(\g/4) \bar{U}^2}
{1+\g[-3\bar{V}'-3\bar{V}+(1/2)\bar{V}'\bar{U}']}\,,\label{syh2}\\
&&\bar{U}''+(3+\zeta) \bar{U}'=6(2+\zeta)\, ,\label{syU2}\\
&&\bar{V}''+(3+\zeta) \bar{V}'=h^{-2} \bar{U}\label{syV}\, .
\ees
In \eqs{syU2}{syV} appears $\zeta=h'/h$. In turn, $h'$ can be computed explicitly from
\eq{syh2}. The resulting expression contains $\bar{V}''$ and $\bar{U}''$, which can  be eliminated through
\eqs{syU2}{syV}. This gives
\be
\zeta= \frac{1}{2(1-3 \gamma\bar{V})}
\[ h^{-2}\Omega'+3\g\(h^{-2}\bar{U}+\bar{U}'\bar{V}'-4\bar{V}'\)\]\, ,
\ee
where $\Omega(x)=\Omega_M e^{-3x}+\Omega_R e^{-4x}$. Then \eqs{syU2}{syV}, with $\zeta$ given by the above expression and $h^2$ given by \eq{syh2}, provide a closed set of second order equations for $\bar{V}$ and $\bar{U}$, whose numerical integration is straightforward.\footnote{As initial conditions we set $U=U'=V=V'=0$, at an initial time $x_{\rm in}$ deep into the RD phase. This choice of initial conditions is a part of the definition of the $\iBox$ operators that enter in \eqs{Udef}{Sdef} as discussed in the previous section.}
The result is shown in Fig.~\ref{fig:UV}. 
Plugging the result back into \eq{syh2} we get $h^2(x)$, and
the effective DE density can then be recovered writing $h^2(x)=\Omega(x)+\rho_{\rm DE}(x)/\rho_0$. The result is shown in the left panel of Fig.~\ref{fig:rhoDE}.  
This plot shows that the effective DE density vanishes deep into the RD phase (the RD-MD transition is at $x_{\rm eq}\simeq -8.1$), and then grows as we enter in the MD phase. It is possible to choose $\gamma$ so to reproduce the observed value today, which is already a non-trivial result. In particular, tuning 
$\gamma$  to the value $\gamma\simeq 0.0089247$ (corresponding to $m\simeq 0.283 H_0$) we get $\ode\simeq 0.6825$ and therefore $\oma\simeq 0.3175$,
which is the value suggested by the {\it Planck} data \cite{Ade:2013zuv} (assuming $\Lambda$CDM, a point to which we will return below). The right panel of  Fig.~\ref{fig:rhoDE}
shows again the function $\rde(z)/\rho_0$, now against the redshift $z$. Observe that $\rho_0$ is the total density today, rather than the total density at redshift $z$. In $\Lambda$CDM $\rde(z)/\rho_0$ remains constant, while in the non-local model it slowly decreases with increasing redshift, indicating a `phantom-like' behaviour. In any case,
$\rde(z)$ quickly becomes negligible with respect to $\rho_{\rm tot}(z)$, which instead grows as $(1+z)^3$ in MD. So, there is no early DE in the nonlocal model.

\begin{figure}[t]
\centering
\includegraphics[width=0.45\columnwidth]{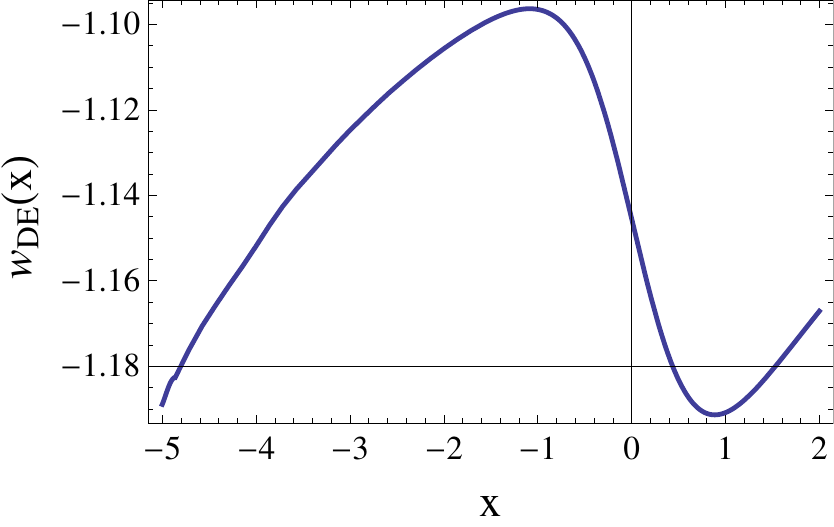}
\includegraphics[width=0.45\columnwidth]{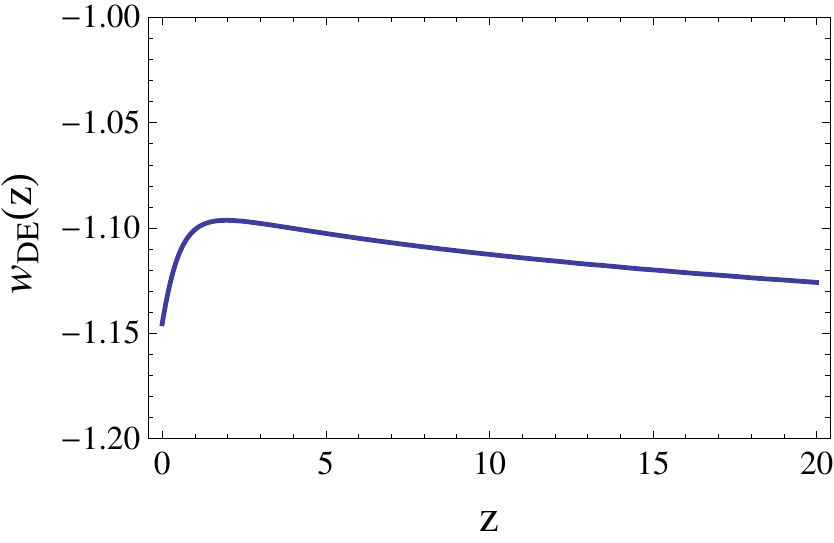}
\caption{\label{fig:wDE} The function $w_{\rm DE}$ against $x$ (left panel) and 
against  the redshift $z=e^{-x}-1$ (right panel), choosing $\gamma$ so that $\oma= 0.3175$. For comparison, the range $0<z<4$ corresponds to $0>x>-1.6$.
}
\end{figure}

\subsection{Prediction for $w_{\rm DE}(z)$}

\begin{figure}[t]
\centering
\includegraphics[width=0.45\columnwidth]{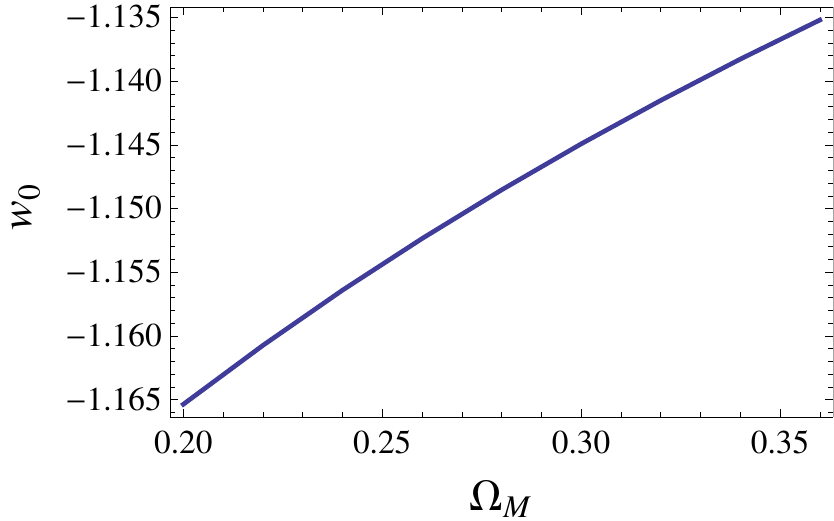}
\includegraphics[width=0.45\columnwidth]{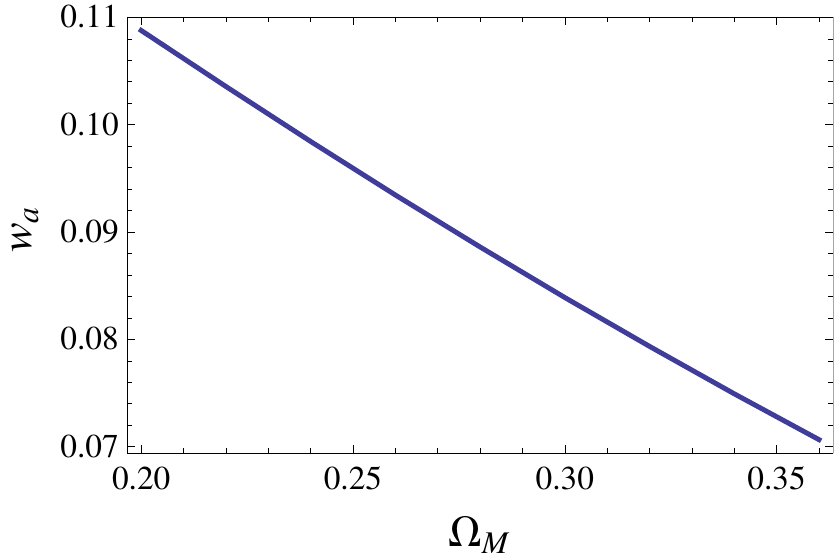}
\caption{\label{fig:wvsOM} Left panel: $w_0$ as a function of $\oma$.
Right panel: $w_a$ as a function of $\oma$.}
\end{figure}

Having fixed  $\gamma$ from the condition of recovering a given value of $\oma$ today, at the level of background evolution there is no more free parameter, and we get a pure prediction for the equation of state (EOS) parameter of dark energy $w_{\rm DE}$, defined from
\be\label{consrho}
{\rho}'_{\rm DE}+3(1+w_{\rm DE})\rho_{\rm DE}=0\, .
\ee
Equivalently, we can define an effective DE pressure $p_{\rm DE}$ from the trace of the $(ii)$ component of the modified Einstein equation (\ref{eqmRiBox2R}), and define 
$w_{\rm DE}=p_{\rm DE}/\rde$. The two definitions are equivalent, upon use of the equations of motion for the auxiliary fields.
The result for $w_{\rm DE}$ is shown in Fig.~\ref{fig:wDE}. Comparing with the standard fit of the form \cite{Chevallier:2000qy,Linder:2002et}
\be\label{ChevLind}
w_{\rm DE}(a)= w_0+(1-a) w_a\, ,
\ee
(where $a(x)=e^x$) in the region $-1<x<0$, one finds 
the best-fit values 
$w_0 = -1.142$ and  $w_a = 0.080$. The value of $\oma$ quoted above has been obtained from the {\it Planck} data assuming the validity
$\Lambda$CDM. The correct value in the nonlocal model should be determined self-consistently with a global fit that takes into account the specific form of the cosmological perturbations in the nonlocal model. It is therefore important to see how the prediction for $w_{\rm DE}$ depends on $\oma$. 
We have  repeated the analysis for different values of $\oma$ in the range $[0.20,0.35]$, adjusting each time $\g$ so as to obtain the desired value of $\oma$.\footnote{We keep fixed
$\oma h_0^2\simeq 0.142$. At the level of background evolution $h_0$ only enters in the determination of the radiation energy density, which we set to $\ora =4.15\times 10^{-5} h_0^{-2}$. We fix $\g$ each time by requiring $h(0)=1$, for a given $\oma$, with a six digit precision. This typically require
fixing $\g$ to 7 digits, so the values of $\g$ are better stored as a table of data, rather than fitted.
Alternatively, one needs a quartic fit 
$\gamma =0.0103959 + 0.00687851 \oma - 0.0598026 \oma^2 + 0.094128 \oma^3 - 
 0.0624636 \oma^4$ to reproduce $\g$ to the necessary precision in the region
 $\oma\in [0.20,0.35]$. Observe also that the value of $\gamma$ has a slight dependence on the value of $x_{\rm in}$ where we start the numerical integration. At the level of seven digits, this dependence becomes negligible if we take $x_{\rm in}\leq -20$.} 
The result for $w_0$ and $w_a$ as a function of $\oma$ are shown in Fig.~\ref{fig:wvsOM}.
In the region  $\oma\in [0.20,0.35]$,
up to the third decimal figure (included), these values are reproduced by the fits 
\bees
w_0&\simeq & -1.2018 + 0.1877\, \oma \, , \label{w0oma}  \\
w_a&\simeq &  \phantom{+}0.1558 - 0.2384\,  \oma\, .
\ees 
Thus, even  varying $\oma$ over the rather broad range $\oma\in [0.20,0.36]$,
$w_0$ remains within the relatively narrow interval $[-1.165,-1.135]$, while
$w_a\in [0.07,0.11]$.
These results fully characterize the model, at the level of the background evolution. 

\subsection{Comparison with SNe Ia data}

Supernova data are mostly sensitive to the background evolution of the cosmological model.
To test whether the background evolution found above is in agreement with distance measurements
of type-Ia supernovae (SNe Ia) we have compared its predictions with the recent joint analysis (``JLA'') of the SDSS-II
and SNLS supernova samples \cite{Betoule:2014frx}. We used the analysis module provided by the supernova collaboration,
varying the cosmological parameters $\oma$ and $H_0$,\footnote{The module marginalises internally over one or two absolute
magnitudes that are degenerate with $H_0$, and indeed we find no constraint on $H_0$ as expected.} as well as the nuisance parameters
$\alpha$ and $\beta$ of the SALT2 light curve model. We assume a spatially flat geometry, we set the radiation
energy density today to $\ora = 4.15 \times 10^{-5}/h_0^2$, and we fix the parameter $\gamma$
of the nonlocal models by requiring that $H(a=1)=H_0$. For the $\Lambda$CDM model we
find parameter constraints in agreement with table 10 of \cite{Betoule:2014frx} although the minimal
$\chi^2$ value returned by the likelihood module lies in between the `stat+sys' and the `stat' values given
in that table.

We find that the ``$R\,\Box^{-2}R$ model" fits the SNe Ia data roughly as well as $\Lambda$CDM, with a minimal $\chi^2$
that is 0.9 higher (which is not significant). However, due to the lower value of the equation of state the model prefers
a slightly higher matter density, $\oma = 0.341 \pm 0.031$ (which, according to \eq{w0oma},
gives $w_0\simeq -1.138$ and $w_a\simeq 0.075$)
compared to $\oma = 0.297 \pm 0.034$ for $\Lambda$CDM. Of course, the best-fit value of $\oma$ for the nonlocal model must eventually be determined through a global fit to SNe, CMB and structure formation.

We also tested the background evolution predicted by the ``$\gmn\iBox R$ model" from \cite{Maggiore:2013mea}.
Not surprisingly the results lie in between the other two models, with $\oma = 0.314 \pm 0.033$. We summarise
the values in table \ref{tab:bgfit}. Based on the very similar goodness of fit and width of the constraints on $\oma$
we conclude that the Bayesian model probabilities are all comparable, with no model being significantly preferred by
the JLA SNe Ia data.

\begin{table}
\begin{center}
\begin{tabular}[tb]{lcc}
\hline
model & $\oma$ & $\chi^2_{\rm min}$ \\
\hline
$\Lambda$CDM  & $0.297 \pm 0.034$ & 695.1 \\
$R\,\Box^{-2}R$  & $0.341 \pm 0.031$ & 696.0 \\
$\gmn\iBox R$     & $0.314 \pm 0.033$ & 695.3 \\
\hline
\end{tabular}
\caption{\label{tab:bgfit} Goodness of fit and matter density today for the nonlocal models and $\Lambda$CDM
when compared to the JLA SNe Ia data. All models fit the SNe Ia data about equally well, but the nonlocal models
prefer a slightly higher matter density.}
\end{center}
\end{table}

For the remainder of this paper we will use for all models the {\it Planck} $\Lambda$CDM best-fit value for the matter
density, $\oma = 0.3175$, in order to compare the perturbations for the same matter abundance and to avoid mixing
the perturbation evolution with effects due to a different $\oma$.

\section{Scalar perturbations}\label{sect:pert}

\subsection{Perturbation equations}

We now study the evolution of scalar perturbations in this model. We work in the  Newtonian gauge, 
\be\label{defPhiPsi}
ds^2 =  -(1+2 \Psi) dt^2 + a^2(t) (1 + 2 \Phi) \delta_{ij} dx^i dx^j\, ,
\ee
(our notations are as in \cite{Dodelson_book,Luca_book}), we use $V=H_0^2S$
and we expand the auxiliary fields as
$U=\bar{U}+\d U$, $V=\bar{V} +\d V$.
Thus, in this model the scalar  perturbations  are described by $\Psi,\Phi,\d U$ and $\d V$.
Let us also recall that, for a generic anisotropic fluid, at first order in perturbation theory we have
\bees
T^0_0 &=& -(\bar{\rho} +\delta \rho),\label{T00} \\
T^0_i &= & (\bar{\rho} + \bar{p})v_i, \label{T0i}\\
T^i_j &= &(\bar{p} + \delta p) \delta^i_j + \Sigma^i_j\, ,\label{Tij}
\ees
where $\bar{\rho}$ and $\bar{p}$ are the unperturbed density and pressure.
The perturbation variables are $\delta \rho, \delta p$, $v_i$, and 
the anisotropic stress tensor $\Sigma^i_j$, which is symmetric  and  traceless, 
$\Sigma^i_i = 0$. The pressure perturbations can be written as $\d p =c_s^2\d \rho$, where $c_s^2$ is the speed of sound of the fluid, and we define as usual $\delta \equiv \delta \rho / \bar{\rho}$ and $\theta \equiv \delta^{ij} \partial_i v_j$. 
On the right-hand side of \eq{eqmRiBox2R} we only put radiation plus non-relativistic matter, so in this case $\Sigma^i_j\simeq0$. However, the general form 
(\ref{T00})--(\ref{Tij}) will be useful in order to recast the nonlocal term in 
\eq{eqmRiBox2R} as the  energy-momentum tensor of an effective fluid.
We can now 
linearize the modified Einstein equations (\ref{eqmRiBox2R}). We perform directly the spatial Fourier transform and we write the time derivatives in terms of $x=\ln a$. We define
$\hat{k}=k/(aH)$,  $\hat{\theta}=\theta/(aH)$ and we use a prime to denote $\pa/\pa x$.
The $(00)$ component of \eq{eqmRiBox2R} gives
\bees
&&\hspace*{-2cm}\( 1 - 3 \gamma \bar{V} \) \( \hat{k}^2 \Phi + 3  \Phi' -3 \Psi  \) + \frac{3 \gamma}{2} \bigg[  - \frac{1}{2 h^2} \bar{U} \delta U + \big( 6 \Psi - 3 \Phi' - \Psi \bar{U}' \big)  \bar{V}' 
 \nn \\
&& + \frac{1}{2} \big( \bar{U}'  \delta V' +  \bar{V}'  \delta U' \big) - 3 \delta V  - 3 \delta V' - \hat{k}^2 \delta V \bigg] = \frac{3}{2 \rho_0 h^2} \bar{\rho} \delta\, .\label{eqlin00}
\ees
The divergence of the $(0i)$ component gives
\be
\( 1 - 3 \gamma \bar{V} \)  \hat{k}^2 (  \Phi' - \Psi) - \frac{3 \gamma \hat{k}^2}{2}
\[ \delta V' -  \bar{V}' \Psi - \delta V + \frac{1}{2} \(  \bar{U}' \delta V +\bar{V}' \delta U \)  \] 
= - \frac{3}{2 \rho_0 h^2}  \hat{\theta} \bar{\rho} ( 1 + w)\, ,\label{eqlin0i}
\ee
The trace of the $(ij)$ component gives
\bees
&&( 1 - 3 \gamma \bar{V} )\[   \Phi'' + (3+\zeta) \Phi' -  \Psi' - (3+2 \zeta) \Psi + \frac{\hat{k}^2}{3}(\Phi + \Psi) \] 
 \nn \\ 
&& - \frac{3 \gamma}{2} \bigg[ \frac{1}{2 h^2} \bar{U} \delta U - 2 \Psi  \bar{V}'' + 
\big[ 2 \Phi' - 2 (2 + \zeta) \Psi -  \Psi' - \Psi  \bar{U}' \big]  \bar{V}' +  \delta V'' + (2+\zeta)  \delta V' \nn \\
&& \hspace{1cm} + \frac{2 \hat{k}^2}{3} \delta V + ( 3 + 2 \zeta ) \delta V + \frac{1}{2} \big(  \bar{U}'  \delta V'  +  \bar{V}'   \delta U' \big) \bigg] = - \frac{3}{2 \rho_0 h^2}  \delta p\, ,
\label{eqlintraceij}
\ees
while, applying the projector $(\n^{-2}\pa_i\pa_j-\frac{1}{3}\d_{ij})$ to the $(ij)$ component to extract the traceless part, we get
\be\label{eqlintransvij}
( 1 - 3 \gamma \bar{V}) \hat{k}^2 (\Psi + \Phi)  - 3 \gamma \hat{k}^2 \delta V = \frac{9}{2 \rho_0 h^2}  e^{2x} \bar{\rho} ( 1 + w) \sigma\, ,
\ee
where $\sigma$ is defined by
\be
\bar{\rho} (1+w) \sigma \equiv \frac{1}{a^2} \frac{\partial^i \partial^j}{\n^2} \Sigma_{ij}. \label{sigma}
\ee
For $\gamma =0$ these four equations reduce to the standard GR result, see e.g. \cite{Luca_book}, as they should. The system of equations is completed linearizing \eqs{BoxU}{BoxS}, which gives
\bees
&& \hspace{-10mm}\delta U'' + (3 + \zeta)  \delta U' + \hat{k}^2 \delta U - 2 \Psi \bar{U}'' - \big[ 2 (3+\zeta) \Psi  +  \Psi' - 3  \Phi' \big]  \bar{U}'  \nn\\
&&  = 2 \hat{k}^2 (\Psi + 2 \Phi) + 6 \big[\Phi'' +(4+\zeta)  \Phi' \big] 
- 6 \big[ \Psi' + 2(2+ \zeta) \Psi \big], \label{eqlinU}\\
&& \hspace{-10mm}\delta V'' + (3 + \zeta) \delta V' + \hat{k}^2 \delta V - 2 \Psi  \bar{V}'' - 
\[ 2 (3+\zeta) \Psi  +  \Psi' - 3  \Phi' \]  \bar{V}' = h^{-2} \delta U\, .\label{eqlinV}
\ees
The energy-momentum tensor on the right-hand side of \eqst{eqlin00}{eqlintransvij} corresponds to the case of a single fluid. In our case we must includes both matter and radiation. Then,  the expressions on the right-hand side of \eqst{eqlin00}{eqlintransvij} are actually given by
\bees
 \bar{\rho} \delta&=&\d\rho_M+\d\rho_R\, ,\\
 \theta \bar{\rho} (1 + w) &=& \theta_M \bar{\rho}_M (1 + w_M)
 +\theta_R \bar{\rho}_R (1 + w_R)= \theta_M \bar{\rho}_M +
(4/3)\theta_R \bar{\rho}_R \, ,\\
\d p&=&\d p_M+\d p_R = c_{s,M}^2\d\rho_M+ c_{s,R}^2\d\rho_R=(1/3)\d\rho_R \, ,\\
c_s^2 \bar{\rho} \delta &=& c_{s,M}^2\bar{\rho}_M \delta_M+
c_{s,R}^2\bar{\rho}_R \delta_R= (1/3) \bar{\rho}_R \delta_R \, ,
\ees
where $\d_M=\d\rho_M/\rho_M$, $\d_R=\d\rho_R/\rho_R$, and we used $w_M=c_{s,M}^2=0$ and $w_R=c_{s,R}^2=1/3$. For matter and radiation, we take $\s=0$ on the right-hand side of \eq{eqlintransvij}.
Using the expressions appropriate to the matter-radiation fluid in \eqst{eqlin00}{eqlintransvij} we  get 
\bees
&&\hspace*{-1.5cm}\( 1 - 3 \gamma \bar{V} \) \( \hat{k}^2 \Phi + 3  \Phi' -3 \Psi  \) + \frac{3 \gamma}{2} \bigg[  - \frac{1}{2 h^2} \bar{U} \delta U + \big( 6 \Psi - 3 \Phi' - \Psi \bar{U}' \big)  \bar{V}' 
 \nn \\
&&\hspace*{-1.5cm} + \frac{1}{2} \big( \bar{U}'  \delta V' +  \bar{V}'  \delta U' \big) - 3 \delta V  - 3 \delta V' - \hat{k}^2 \delta V \bigg] =  \frac{3}{2 h^2} \big(  \Omega_{\textsc{R}} e^{-4x} \delta_{\textsc{R}} + \Omega_{\textsc{M}} e^{-3x} \delta_{\textsc{M}} \big)  , \label{MEE1x}\\
&&\hspace*{-1.5cm}\( 1 - 3 \gamma \bar{V} \)  \hat{k}^2 (  \Phi' - \Psi) - \frac{3 \gamma \hat{k}^2}{2}
\[ \delta V' -  \bar{V}' \Psi - \delta V + \frac{1}{2} \(  \bar{U}' \delta V +\bar{V}' \delta U \)  \] \nn\\
&&\hspace*{-1.5cm}=
 - \frac{3}{2 h^2} \bigg(  \frac{4}{3} \Omega_{\textsc{R}} e^{-4x} \hat{\theta}_{\textsc{R}} +  \Omega_{\textsc{M}} e^{-3x} \hat{\theta}_{\textsc{M}} \bigg), \label{MEE2x}\\
&&\hspace*{-1.5cm}( 1 - 3 \gamma \bar{V} )\[   \Phi'' + (3+\zeta) \Phi' -  \Psi' - (3+2 \zeta) \Psi + \frac{\hat{k}^2}{3}(\Phi + \Psi) \] 
 \nn \\ 
&&\hspace*{-1.5cm} - \frac{3 \gamma}{2} \bigg\{ \frac{1}{2 h^2} \bar{U} \delta U - 2 \Psi  \bar{V}'' + 
\[ 2 \Phi' - 2 (2 + \zeta) \Psi -  \Psi' - \Psi  \bar{U}' \]  \bar{V}' +  \delta V'' + (2+\zeta)  \delta V' \nn \\
&&  + \frac{2 \hat{k}^2}{3} \delta V + ( 3 + 2 \zeta ) \delta V + \frac{1}{2} \big(  \bar{U}'  \delta V'  +  \bar{V}'   \delta U' \big) \bigg\}
= - \frac{1}{2 h^2}  \Omega_{\textsc{R}} e^{-4x} \delta_{\textsc{R}}\, ,\label{MEE3x}\\
&&\hspace*{-1.5cm}
( 1 - 3 \gamma \bar{V})  (\Psi + \Phi)  - 3 \gamma \delta V =0\, .\label{MEE4x}
\ees
As usual, it is convenient to write also the equations derived from the linearization of  energy-momentum conservation $\n^{\mu}\Tmn=0$, even if they are not independent from 
\eqst{eqlin00}{eqlintransvij}. For a single generic fluid the linearization of the $\nu=0$ component gives
\be\label{conspert0}
\delta' = - (3  \Phi' + \hat{\theta}) (1+w) - 3 \delta (c_s^2 -w)\, ,
\ee
while, applying the divergence to the $\nu=i$ equation, we get
\be\label{consperti}
 \hat{\theta}' = - \( 2-3w + \zeta + \frac{ w'}{1+w} \) \hat{\theta} 
 + \hat{k}^2 \( \Psi + \sigma + \frac{c_s^2}{1+w} \delta \)\, .
\ee
Observe that these equations are independent of the specific DE content of a theory, since they just express the conservation of $\Tmn$. For the matter-radiation fluid,  with no energy exchange among them, we have the usual equations~\cite{Luca_book}
\begin{align}
\delta_M' &= - (3  \Phi' + \hat{\theta}_M),\label{dM1} \\
\hat{\theta}'_M &= - ( 2 + \zeta) \hat{\theta}_M + \hat{k}^2 \Psi\, ,\label{dtheta1}\\
\delta_R' &= - \frac{4}{3} (3  \Phi' + \hat{\theta}_R), \\
 \hat{\theta}'_R &= - (1 + \zeta)\hat{\theta}_R + \hat{k}^2 \( \Psi + \frac{\delta_R}{4} \). \label{dtheta2}
\end{align}
In particular, taking the derivative of \eq{dM1} and using \eq{dtheta1}, we get 
\be\label{dM2}
\d''_M+(2+\zeta)\d'_M=-3\[ \Phi''+(2+\zeta)\Phi'\] -\hat{k}^2 \Psi\, .
\ee

\subsection{Effective fluid description of the nonlocal theory}

In Sect.~\ref{sect:back} we have seen how, at the level of background evolution, the nonlocal term in \eq{eqmRiBox2R} acts effectively as a fluid with an energy density
$\rde$ and a pressure $p_{\rm DE}=w_{\rm DE}\rde$. The same effective
fluid description can be applied to  the perturbations induced by the nonlocal terms. 
The four linearized Einstein equations \eqst{MEE1x}{MEE4x} can in fact  be rewritten as
\bees
&&\hat{k}^2 \Phi + 3( \Phi' - \Psi) = \frac{4 \pi G}{H^2} \sum_i \delta \rho_i , \label{EE1} \\
&&\hat{k}^2 \big(  \Phi'  - \Psi \big) = - \frac{4 \pi G}{H^2}  \sum_i \bar{\rho}_i (1+w_i) \hat{\theta}_i , \label{EE2} \\
&&\hat{k}^2 (\Psi + \Phi) = \frac{12 \pi G e^{2x}}{H^2} \bar{\rho}_{\rm DE} (1+w_{\rm DE}) \sigma_{\rm DE}, \label{EE3} \\
&&\Phi'' + (3+\zeta) \Phi' -  \Psi' - (3+2\zeta) \Psi + \frac{\hat{k}^2}{3} (\Phi + \Psi) = - \frac{4 \pi G}{H^2} \sum_i \delta p_i\, , \label{EE4}
\ees
where the sums over $i$ run over  radiation, matter and dark energy, and we have defined
\bees
\delta \rde&=&  \rho_0   \gamma h^2
\bigg[  2 \bar{V} \( \hat{k}^2 \Phi + 3 \(  \Phi' - \Psi \) \) + \frac{1}{2 h^2} \bar{U} \delta U - 
\big( 6 \Psi - 3  \Phi' - \Psi \bar{U}' \big) \bar{V}'  \nn \\
&& - \frac{1}{2} \big( \bar{U}'  \delta V' +  \bar{V}'  \delta U' \big) + 3 \delta V
 + 3  \delta V' + \hat{k}^2 \delta V  \bigg],\label{drhoDE}\\
 \rde \( 1+w_{\scDE}\)\hat{\theta}_{\scDE} &=& - \rho_0  \gamma h^2\hat{k}^2 
 \[ 2 \bar{V} ( \Phi' - \Psi) +  \delta V' -  \bar{V}' \Psi - \delta V 
 + \frac{1}{2} \(  \bar{U}' \delta V + \bar{V}' \delta U \) \],\nn\\
 \\
\rde  (1+ w_\scDE) \sigma_\scDE &=&  \frac{2}{3}\rho_0 h^2 \hat{k}^2 \gamma
e^{-2x} \[ \bar{V} \( \Phi + \Psi \) + \delta V \]\, ,
\ees
\bees
\delta p_{\scDE} &=& -  \rho_0\gamma h^2 
\bigg[ 2 \bar{V} \(  \Phi'' + (3+\zeta) \Phi' -  \Psi' - (3+2 \zeta) \Psi 
+ \frac{\hat{k}^2}{3}(\Phi + \Psi) \) \nn \\
&&+ \frac{1}{2 h^2} \bar{U} \delta U - 2 \Psi \bar{V}'' + 
\bigg( 2 \Phi' - 2 (2 + \zeta) \Psi - \Psi' - \Psi  \bar{U}' \bigg) \bar{V}' \nn \\
& &+  \delta V'' + (2+\zeta)  \delta V' + \frac{2 \hat{k}^2}{3} \delta V + ( 3 + 2 \zeta ) \delta V + \frac{1}{2} \big(  \bar{U}'  \delta V'  +\bar{V}'   \delta U' \big) \bigg].\label{dpDE}
\ees
The EOS parameter $w_{\rm DE}$ is defined in \eq{consrho}, and we have made  use of the equations for the background, \eq{syh2}. From these quantities we can form the combination
\be
\d_{\rm DE}=\frac{\d\rde}{\rde}\, ,
\ee
and we can define an effective speed of sound for the DE perturbations,\footnote{We use a hat to stress that this is not the usual rest-frame speed of sound. At any rate, for large $\hat{k}$ the difference between the two definitions vanishes.}
\be\label{csDE}
\hat{c}^2_{s,\scDE}(x) \equiv \frac{\delta p_{\scDE}(x)}{\delta \rho_{\scDE}(x)}\, .
\ee
The above quantities completely characterize the DE perturbations as an effective fluid. It is worth stressing  that   the $R\,\Box^{-2}R$ model is remarkably predictive. In terms of a single  free parameter $\gamma$ that replaces $\ola$ in $\Lambda$CDM, the model fully predicts the function $w_{\rm DE}(x)$ as a function of $x=\ln a$ (or, equivalently, of the redshift $z$, with $z=e^{-x}-1$), which characterizes the background evolution, as well as the functions $\d_{\rm DE}(a;k)$, 
$\hat{\theta}_{\rm DE}(a;k)$, $\s_{\rm DE}(a;k)$ and $\hat{c}^2_{s,\scDE}(a;k)$, which fully characterize the perturbations. Similar considerations hold for
the $\gmn\iBox R$ model for which a similar treatment is possible, as we discuss in
app.~\ref{app:gmnR}, see also
\cite{Nesseris:2014mea}.

\subsection{Indicators of deviations from GR}

It can be useful to extract from the above equations some indicators that give a simple way to estimate the deviations of the results from those obtained in GR (see e.g.\ \cite{Kunz:2012aw} for a short review). One such indicator is obtained combining \eqref{MEE1x} and \eqref{MEE2x} to get a modified Poisson equation, 
\begin{align}
(1-3\gamma\bar{V})\hat{k}^2 \Phi &= \frac{3}{2 h^2} \bigg[ \Omega_R e^{-4x} \bigg( \delta_R + \frac{4}{\hat{k}^2} \hat{\theta}_R \bigg) + \Omega_M e^{-3x} \bigg( \delta_M + \frac{3}{\hat{k}^2} \hat{\theta}_M \bigg) \bigg] \nn \\
&- \frac{3 \gamma}{2} \bigg[  - (\hat{k}^2 + 6 ) \delta V - \frac{1}{2 h^2} \bar{U} \delta U + \big( 3 \Psi - 3  \Phi' - \Psi \bar{U}' \big)  \bar{V}' \nn \\
& + \frac{1}{2}   \bar{U}' (  \delta V' + 3 \delta V ) + \frac{1}{2}\bar{V}' (  \delta U' + 3 \delta U)  \bigg]\label{modPois1}
\end{align}
We can then define a function $G_{\rm eff}(x;k)$ through\footnote{Recall that $x=\ln a(t)$ is the time evolution variable while $k=|{\bf k}|$ is the modulus of the comoving spatial momentum; since we work directly in momentum space, all functions $\Psi,\Phi,\delta U$ and $\d V$ are functions of $x$ and $k$, even if we do not write their dependence explicitly.}
\be\label{defGeff}
\frac{G_{\rm eff}(x;k)}{G} \equiv  \frac{1}{1-3\gamma\bar{V}(x)}\, \[ 1 - \frac{P(x;k)}{R(x;k)}\]\, ,
\ee
(this function is also occasionally called $Q$, e.g.\ in \cite{Amendola:2007rr}) where
\bees
P(x;k) &\equiv& \frac{3 \gamma}{2} \bigg[  - (\hat{k}^2 + 6 ) \delta V - \frac{1}{2 h^2} \bar{U} \delta U + \big( 3 \Psi - 3 \Phi' - \Psi \bar{U}' \big)  \bar{V}' \nn \\
&& + \frac{1}{2}  \bar{U}' (  \delta V' + 3 \delta V ) +\frac{1}{2}  \bar{V}' (  \delta U' + 3 \delta U)  \bigg],\label{defP}
\ees 
\be\label{defR}
R(x;k) \equiv \frac{3}{2 h^2} \bigg[ \Omega_R e^{-4x} \bigg( \delta_R + \frac{4}{\hat{k}^2} \hat{\theta}_R \bigg) + \Omega_M e^{-3x} \bigg( \delta_M + \frac{3}{\hat{k}^2} \hat{\theta}_M \bigg) \bigg]\, .
\ee
Then the modified Poisson equation (\ref{modPois1}) can be rewritten as
\be\label{PhiGeff}
k^2\Phi =4\pi G_{\rm eff}(x;k) a^2 \rho_0
\[ \Omega_R e^{-4x} \bigg( \delta_R + \frac{4}{\hat{k}^2} \hat{\theta}_R \bigg) + \Omega_M e^{-3x} \bigg( \delta_M + \frac{3}{\hat{k}^2} \hat{\theta}_M \bigg) \]\, .
\ee
This shows that $G_{\rm eff}(x;k) $ plays the role of an effective time-dependent 
gravitational constant, which also depends on the mode $k$. Together with $G_{\rm eff}$,  a second useful 
indicator is \cite{Amendola:2007rr,Park:2012cp,Dodelson:2013sma}
\be\label{defeta}
\eta(x;k) =\frac{\Phi+\Psi}{\Phi}\, .
\ee
Alternatively, two useful quantities are the functions $\mu(x;k)$ \cite{Daniel:2010ky} and $\Sigma(x;k)$
\cite{Amendola:2007rr} which are defined through\footnote{Here we will define these functions with respect to the perturbations
in $\Lambda$CDM as this allows us to use the indicator functions to compare directly the perturbation evolution in these two models. In many references, for example in \cite{Amendola:2007rr}, the indicator functions instead represent the additional contribution to the metric perturbations from a dark energy fluid or a modification of gravity. This is not important for $G_{\rm eff}$ and $\eta$, but it changes the interpretation of $\mu$ and $\Sigma$ somewhat, as discussed in more detail in the text. Also, in the literature the quantity that we call $1+\mu$ is sometimes denoted by $\mu$, and similarly our $1+\Sigma$ is sometimes denoted by $\Sigma$. Our definitions are such that in GR $\mu=\Sigma=0$, while with the other definition GR corresponds to $\mu=\Sigma=1$.}
\bees
\Psi&=&[1+\mu(x;k)]\Psi_{\rm GR}\, ,\label{defmu}\\
\Psi-\Phi&=&[1+\Sigma(x;k)] (\Psi-\Phi)_{\rm GR}\label{defSigma}\, ,
\ees
where the subscript denotes the same quantities computed in GR, assuming a $\Lambda$CDM model with the same value of $\oma$ as the modified gravity model. The advantage of this  parametrization is that it neatly separates  the modifications to the motion of non-relativistic particles, which is described by $\mu$, from the modification to light propagation, which is encoded in 
$\Sigma$. These functions have also  been used recently in \cite{Dodelson:2013sma} to compare the Deser-Woodard model with the cosmological data and  in 
\cite{Zhao:2010dz,Daniel:2010ky,Song:2010fg,Simpson:2012ra} 
to put constraints on generic deviations from GR. For modes well inside the horizon, to leading order  
\eq{dM2} becomes
\be\label{dM2b}
\d_M''+(2+\zeta)\d_M'= -\hat{k}^2 [1+\mu(x;k)]\Psi_{\rm GR}\, .
\ee
We now use $\Psi_{\rm GR}=-\Phi_{\rm GR}$ and
$k^2\Phi_{\rm GR}=4\pi G a^2\rho_M(\d_M)_{\rm GR}$, where
$(\d_M)_{\rm GR}$ are the matter density perturbation in general relativity, assuming $\Lambda$CDM. Observe that these are in general different from the density perturbations $\d_M$ in the nonlocal model.
Recalling that $\hat{k}=k/(aH)$, one finds
\be\label{dM3}
\d_M''+(2+\zeta)\d_M' -\frac{3}{2} \[ (1+\mu)\frac{(\d_M)_{\rm GR} }{\d_M}\]\, \frac{\oma\d_M}{a^3 h^2(x)}=0\, ,
\ee
or, using $a$ instead of $x$,
\be\label{dmpp}
\frac{d^2\d_M}{da^2}+\(\frac{3}{a}+\frac{d\ln H}{da}\)\frac{d\d_M}{da} 
-\frac{3}{2}  \[ (1+\mu)\frac{(\d_M)_{\rm GR} }{\d_M}\]\,
\frac{\oma\d_M}{a^5 h^2(a)}=0\, .
\ee
The relation between the sets $(G_{\rm eff}/G,\eta)$ and $(\mu,\Sigma)$ can be obtained as follows.
From \eq{defeta} we have $\Psi=-(1-\eta)\Phi$.  From the definition of $G_{\rm eff}$ however one cannot conclude simply that   $\Phi=(G_{\rm eff}/G)\Phi_{\rm GR}$, because the quantities $\d_M$ and $\hat{\theta}_M$ that appear
in \eq{PhiGeff} are the ones in the nonlocal model and, again, are in general different from the corresponding quantities in a $\Lambda$CDM with the same value of $\oma$ (the same is in principle true for  $\d_R$ and $\hat{\theta}_R$, which however play no role in structure formation during MD). For $\hat{k}^2\gg 1$ the terms $\hat{\theta}_M$ is sub-leading, so only the difference in $\d_M$ is relevant.
Then, 
\be
\Phi= \frac{G_{\rm eff}}{G}\frac{\d_M}{(\d_M)_{\rm GR} }\Phi_{\rm GR}
=-\frac{G_{\rm eff}}{G}\frac{\d_M}{(\d_M)_{\rm GR} }\Psi_{\rm GR}\, .
\ee
This  gives 
\be\label{muetaG}
1+\mu=(1-\eta)\frac{G_{\rm eff}}{G} \frac{\d_M}{(\d_M)_{\rm GR} }\, ,
\ee
Thus, the quantity
$[ (1+\mu)(\d_M)_{\rm GR} /\d_M]$ that appears in \eq{dM3}, and characterizes the difference in the growth of structure between a modified gravity model and $\Lambda$CDM, is equal to 
$(1-\eta)G_{\rm eff}/G$.
Similarly one finds that
\be\label{SetaG}
1+\Sigma= \(1-\frac{\eta}{2}\)\frac{G_{\rm eff}}{G} \frac{\d_M}{(\d_M)_{\rm GR} }\, .
\ee
Observe that \eq{muetaG} reduces to the expression given in \cite{Dodelson:2013sma} only if we set
$\d_M/ (\d_M)_{\rm GR}=1$. This can be useful for order-of-magnitude estimate, but not for accurate quantitative computations.

\section{The sub-horizon limit and structure formation}\label{sect:analytic}

The Fourier modes relevant to the linear regime of structure formation correspond approximately to
\be\label{rangekH0}
30\, \lsim\, k/H_0\, \lsim\, 300\, ,
\ee
and remain well inside the horizon, $k/H(z)\gg 1$, in the redshift range $z\,\lsim\, 2$ relevant to the present observations of redshift space distortion. For these modes we can therefore keep only the leading terms in the limit  $\hat{k}=k/(aH)\gg 1$ in \eqs{eqlinU}{eqlinV}, which then give
\begin{figure}[t]
\centering
\includegraphics[width=0.45\columnwidth]{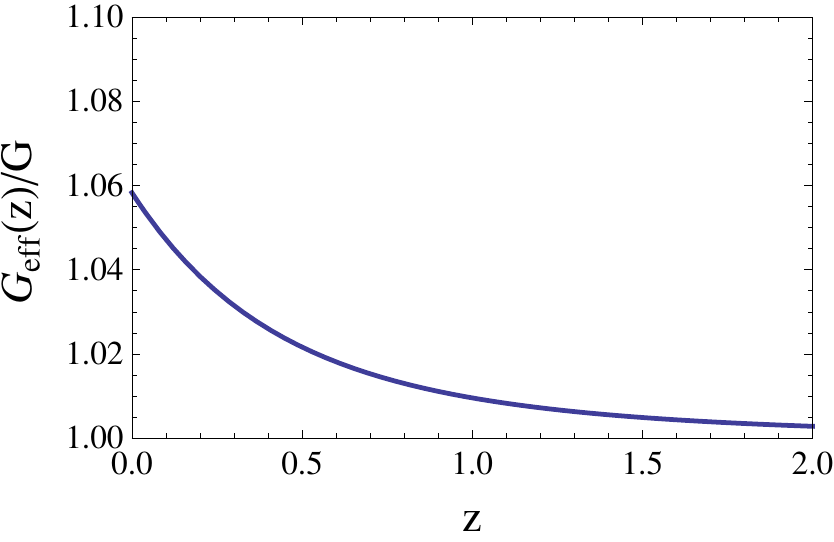}
\caption{\label{fig:Geff}  Left panel: $G_{\rm eff}/G$ as a function of the redshift $z$, for sub-horizon modes.
}
\end{figure}

\be\label{UVlargek}
\d U=2(\Psi+2\Phi)\, ,\qquad \d V={\cal O}\(\frac{1}{\hat{k}^2}\)\Psi\, .
\ee
This shows that the term $P(x;k)$ defined in \eq{defP} is ${\cal O}(1)$ with respect to the large parameter $\hat{k}$, and linear in the perturbation,  i.e. it is overall ${\cal O}(\Phi)$. In contrast, the left-hand side of the Poisson equation (\ref{modPois1}) is ${\cal O}(\hat{k}^2\Phi)$. Thus, in the right-hand side of (\ref{modPois1}), $P$ is subdominant in the large $\hat{k}$ limit, and the left-hand side must be balanced uniquely by $R$, i.e. $R={\cal O}(\hat{k}^2\Phi)$. Then, from
\eq{defGeff} it follows that
\be\label{Geff1suk2}
\frac{G_{\rm eff}(x;k)}{G} =  \frac{1}{1-3\gamma\bar{V}(x)}\, \[ 1 +{\cal O}\(\frac{1}{\hat{k}^2}\)\]\, .
\ee
This shows that, in the sub-horizon limit, $G_{\rm eff}(x;k)$ becomes independent of $k$. We see from Fig.~\ref{fig:UV} that $\bar{V}(x)$ grows with $x$. At $x=0$, $\bar{V}(0)\simeq 2.06$ while $\gamma\simeq 0.0089247$, so $3\gamma\bar{V}(0)\simeq 0.055$ is still much smaller than one, and today
\be\label{Gefflargek}
\frac{G_{\rm eff}(x=0, k\gg H_0)}{G}\simeq 1.0583\, .
\ee
In   Fig.~\ref{fig:Geff} we plot $G_{\rm eff}/G$ as a function of the redshift $z=e^{-x}-1$,  for these sub-horizon modes. We see that in the recent epoch, and for sub-horizon modes, the effective Newton constant in the nonlocal theory is larger than $G$ by  a few percent. This is different from what happens in the $\gmn\iBox R$ model, where instead $G_{\rm eff}/G=1+{\cal O}(1/\hat{k}^2)$, see \cite{Nesseris:2014mea} and App.~\ref{app:gmnR}.
We can next estimate $\eta$ in the sub-horizon limit. From \eq{MEE4x} we see that $\Psi+\Phi$ is of order $\gamma \d V$ and therefore ${\cal O}(\gamma\Phi/\hat{k}^2)$. Therefore
$\eta ={\cal O}(\gamma/\hat{k}^2)$, so it is parametrically of order $1/\hat{k}^2$ (and is further suppressed by the numerical factor $\gamma \simeq 10^{-2}$). 

These results  allow us to get a first understanding of why this  nonlocal model does not have difficulties in explaining data on structure formation, at the present level of accuracy of the data. Indeed, the analysis presented in  \cite{Simpson:2012ra} using a combination of  lensing, redshift space distortion, $H_0$ and high-$l$ WMAP7 data gives,  for  the deviation of $\Psi$  from GR the result, the value
\be\label{DPsiSimpson}
\frac{\Delta\Psi}{\Psi}=0.05\pm 0.25\, ,
\ee
(at 68\% c.l.), and the result is mostly sensitive to modified gravity models at redshift $z\simeq 0.5$. The details of the analysis depend on a number of assumptions on the background evolution and on the dependence on $\mu$  on $a$.\footnote{In particular,  the value (\ref{DPsiSimpson}) has  
been obtained  in   \cite{Simpson:2012ra} assuming either $\Lambda$CDM or 
wCDM for the background, and assuming   a functional form
$\mu(a)=\mu_0\rde(a)/\rde(0)=\mu_0a^{-3(1+w_{\rm DE})}$. This is  different from  the behavior  in our model, that, as we will see in the next section,  rather predicts $\mu(a)\simeq \mu_sa^s$ with 
$s\simeq 2$.
Furthermore, in \cite{Simpson:2012ra}
$\mu$ is also assumed to be  independent of $k$ on sub-horizon scale, which is indeed the case for our model.} However, independently of the details, it is clear that a viable modified gravity model cannot predict 
a value of $(1+\mu)$ and of $G_{\rm eff}/G$
at these redshifts  much in excess of $1.25$. 
From  Fig.~\ref{fig:Geff} we see that in our nonlocal model, for subhorizon modes,
$G_{\rm eff}(z=0.5)/G \simeq 1.02$. The numerical integration of the next section will  confirm this result,
while for  $\mu$ we will find  $\mu(z=0.5)\simeq 0.04$.
These numbers are comfortably  within the 68\% c.l. limits given in \eq{DPsiSimpson}.
For comparison, the nonlocal model proposed by Deser and Woodard~\cite{Deser:2007jk,Deser:2013uya,Woodard:2014iga} predicts a value of $\mu(z)$ with $\mu(z=0.5)\simeq 0.60$, and is therefore ruled out with great statistical significance~\cite{Dodelson:2013sma}.

We conclude this section stressing that $G_{\rm eff}$ only plays the role of an effective Newton constant for cosmological perturbations over a FRW background. As shown in \cite{Kehagias:2014sda,Maggiore:2014sia}, for static spherically symmetric configurations the corrections to the \Sch solution of GR are
$1+{\cal O}(m^2r^2)$. 
Therefore, taking $m\sim H_0$, the deviations from GR are totally negligible at distances $r$ of the order of the solar system (or, more generally, whenever $r\ll H_0^{-1}$), so the  nonlocal theory passes all solar system constraints. 
Of course there is no contradiction between the fact that static solution at $mr\ll 1$ is governed by $G$ while cosmological perturbations by $G_{\rm eff}$. The static solution of the nonlocal theory at distance $r$ depends on the combination $mr$ and reduces to the  \Sch solution of GR for $mr\ll 1$, while it deviates from it for $mr={\cal O}(1)$. In a homogeneous FRW there is no generic distance scale $r$, and  the relevant scale in our nonlocal model is rather provided  by the Ricci scalar $R$. In RD we simply have $R=0$ so the nonlocal term in ineffective while, in MD, $R={\cal O}(H^2)$ and the relevant lengthscale becomes $H^{-1}$. Thus, after RD the corrections to the cosmological evolutions are 
${\cal O}(m^2/H^2)$ and today, when $H=H_0$, they are no longer parametrically small, since also $m={\cal O}(H_0)$. We 
see however that the corrections to $G_{\rm eff}$ today are still numerically small, of the order of a few percent. In a sense, this is due to a sort of delayed response induced by the $\iBox$ operator. The function $U=-\iBox R$ vanishes in RD, where $R=0$, and only starts to grow after we enter the MD phase. In turn, $V=H_0^2 S=-H_0^2\iBox U$ is sourced by $U$ and its background value $\bar{V}$ only begins to grow when $\bar{U}$ is already large. This hierarchy is clearly seen in fig.~\ref{fig:UV}. Since, for sub-horizon modes, $G_{\rm eff}/G$ only depends on $\bar{V}$, in the end 
$G_{\rm eff}/G$ today is still numerically quite close to one.

\section{Numerical results}\label{sect:numeric}

We now present the results obtained from the numerical integration of the perturbation equations.
We integrate the equations using the initial conditions expanded up to second order \cite{Ruth_book},\footnote{We integrate the system using \eqref{eqlinU}, \eqref{eqlinV}, \eqref{dM1}-\eqref{dtheta2} and the $(ij)$ component of the Einstein equation \eqref{MEE3x} that closes the system. Using the second order version of the initial conditions allows one to satisfy the Poisson equation up to a relative different of one part in $10^9$ between the left and right hand sides at initial integration time $x_{\rm in} = -15$. Furthermore we check  the numerical accuracy by verifying that the Poisson equation is satisfied at any integration time. We find that this is indeed the case up to a relative different of one part in $10^8$ for $\kappa = 0.1$ and $10^6$ for $\kappa = 5$. In contrast, we have found that closing the system of equations using the $(0i)$ component of the Einstein equations is numerically much less reliable, while closing the latter with the $(00)$ one leads to the same results.} 
\begin{align}
\Phi(a_{\rm in},k) &= -\Psi(a_{\rm in},k) = A(k) \frac{3 \sqrt{3}}{\hat{k}_{\rm in}}j_1\big( \hat{k}_{\rm in}/\sqrt{3}\big)\nn\\
& \simeq A(k) \bigg( 1 - \frac{\hat{k}_{\rm in}^2}{30} + \frac{\hat{k}_{\rm in}^4}{2520} + \mathcal{O}(k^6) \bigg) , \label{initcond1}\\
\delta_R(a_{\rm in},k) &= \frac{4}{3} \delta_M(a_{\rm in},k) = \frac{6 A(k)}{\hat{k}_{\rm in}^3} 
\[ \hat{k}_{\rm in}(6-\hat{k}_{\rm in}^2) \cos \big( \hat{k}_{\rm in}/\sqrt{3} \big) + 2 \sqrt{3} \big( \hat{k}_{\rm in}^2 - 3) \sin \big( \hat{k}_{\rm in}/ \sqrt{3} \big) \]\nn \\
&\simeq A(k) \bigg( 2 + \frac{7 \hat{k}_{\rm in}^2}{15} - \frac{23 \hat{k}_{\rm in}^4}{1260} + \mathcal{O}(k^6) \bigg), \\
\theta_M(a_{\rm in},k) &= \theta_R(a_{\rm in},k) = - \frac{3 A(k)}{2 \hat{k}_{\rm in}}
 \[ 6 \hat{k}_{\rm in} \cos \big( \hat{k}_{\rm in}/\sqrt{3} \big) + \sqrt{3}( \hat{k}_{\rm in}^2 - 6) \sin \big( \hat{k}_{\rm in}/\sqrt{3} \big) \] \nn\\
& \simeq A(k) \bigg( - \frac{\hat{k}_{\rm in}^2}{2} + \frac{\hat{k}_{\rm in}^4}{20} + \mathcal{O}(k^6) \bigg).\label{initcond4}
\end{align}
with $\hat{k}_{\rm in} \equiv \hat{k}(x_{\rm in}) = k/[e^{x_{\rm in}} h(x_{\rm in})]$ and 
$A^2(k)=(50 \pi^2/9k^3) (k/H_0)^{n_s-1}\d_H^2$.\footnote{Observe that $\d (t,\vx)$ is dimensionless, while  $\d (t,\vk)$ is defined, as usual, as $\d (t,\vk)=V^{-1/2}\int_V d^3x\, e^{-i\vk{\bf\cdot}\vx}\d (t,\vx)$, where $V$ is a large  spatial volume, so it has dimensions $k^{-3/2}$. With this definition, the relation to the power spectrum is
$\langle \d (t,\vk_1)\d (t,\vk_2)\rangle=P(t,k) (2\pi)^3 V^{-1}\d^{(3)}(\vk_1-\vk_2)$ and therefore
$\langle \d^2 (t,\vk)\rangle=P(t,k)$. The same $V^{-1/2}$ factor appears in the definition of
$\Psi(t,k)$, etc.
We will then evolve and  plot the dimensionless quantities 
$k^{3/2}\d (t,\vk)$, $k^{3/2}\Psi (t,\vk)$, etc.}
In this work,
for the spectral index $n_s$ and the amplitude of the gravitational potential $\d_H$ we take the $\Lambda$CDM values
$n_s \simeq 0.96$ and $\delta_H^2 \simeq 3.2 \times 10^{-10}$. Of  course, in a full analysis in which the perturbations are evolved through a Boltzmann code, also the values of $n_s$ and $\d_H$ in our model will have to be determined self-consistently from a global fit to the data.

\begin{figure}[t]
\centering
\includegraphics[width=0.45\columnwidth]{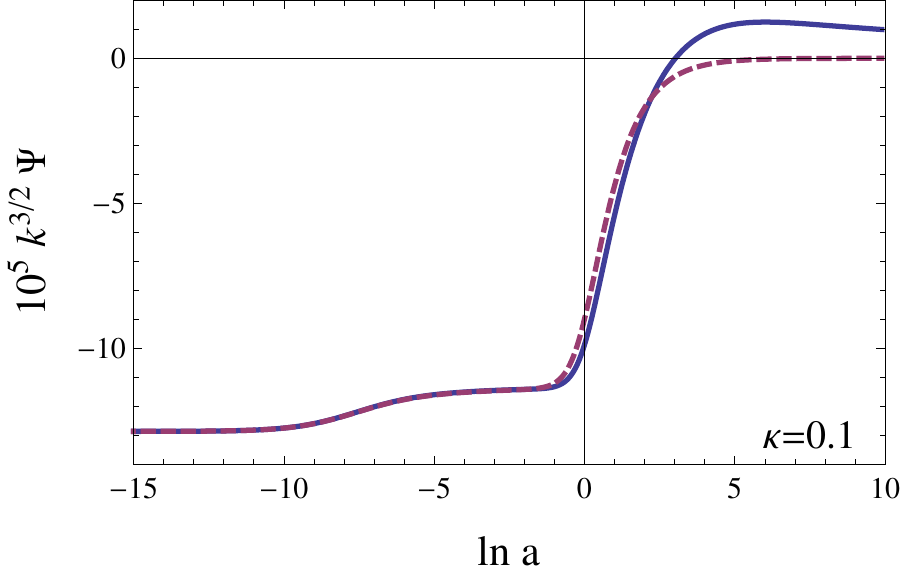}
\includegraphics[width=0.45\columnwidth]{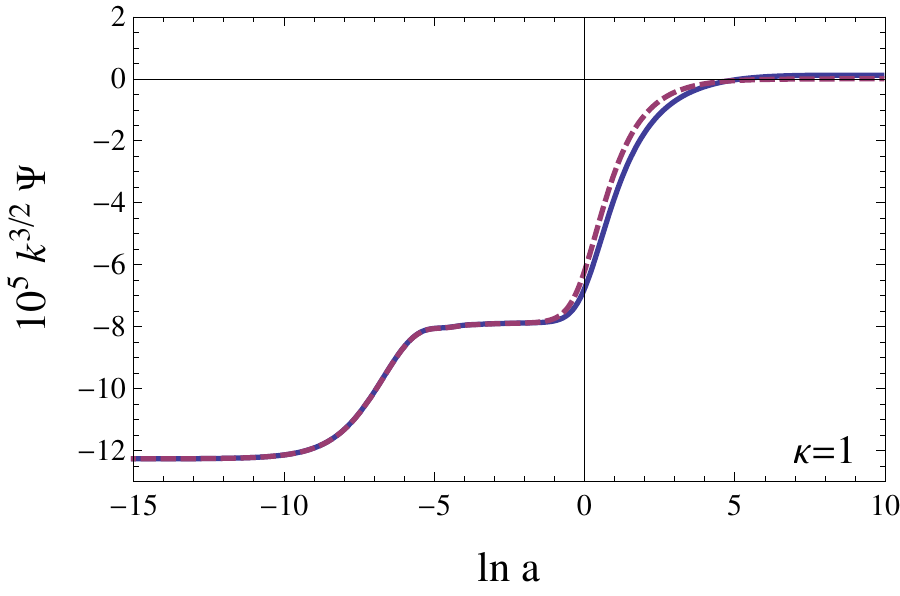}
\includegraphics[width=0.45\columnwidth]{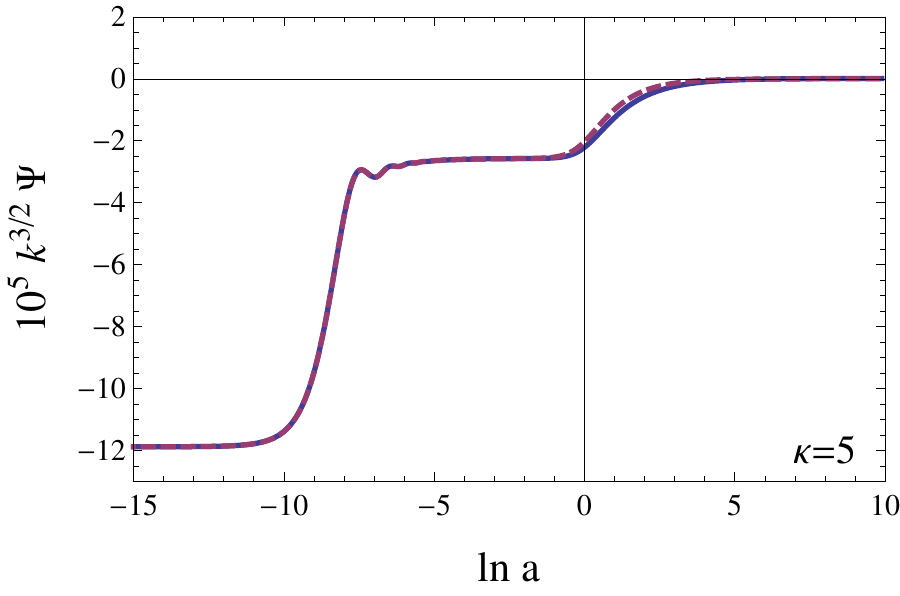}
\caption{\label{fig:F1F2} $k^{3/2}\Psi(a;k)$  from the $R\, \Box^{-2} R$ model (blue solid line) and from $\Lambda$CDM (purple dashed line), as a function of $x=\ln a(t)$, for  
$\kappa = 0.1$ (left upper panel), $\kappa=1$ (right upper panel), $\kappa=5$ (lower panel).
Observe that, on the vertical axis, we plot $10^5k^{3/2}\Psi(a;k)$.
}
\end{figure}

\begin{figure}[t]
\centering
\includegraphics[width=0.45\columnwidth]{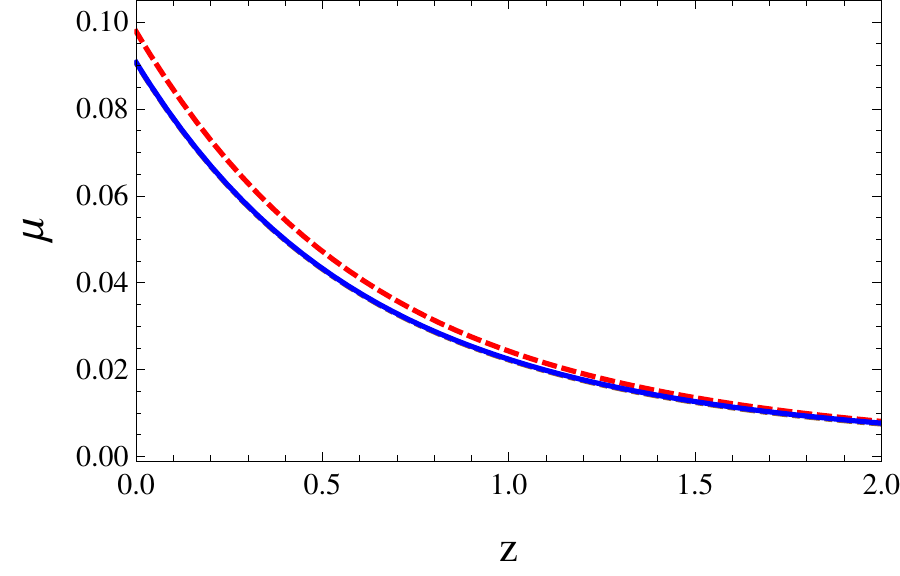}
\includegraphics[width=0.45\columnwidth]{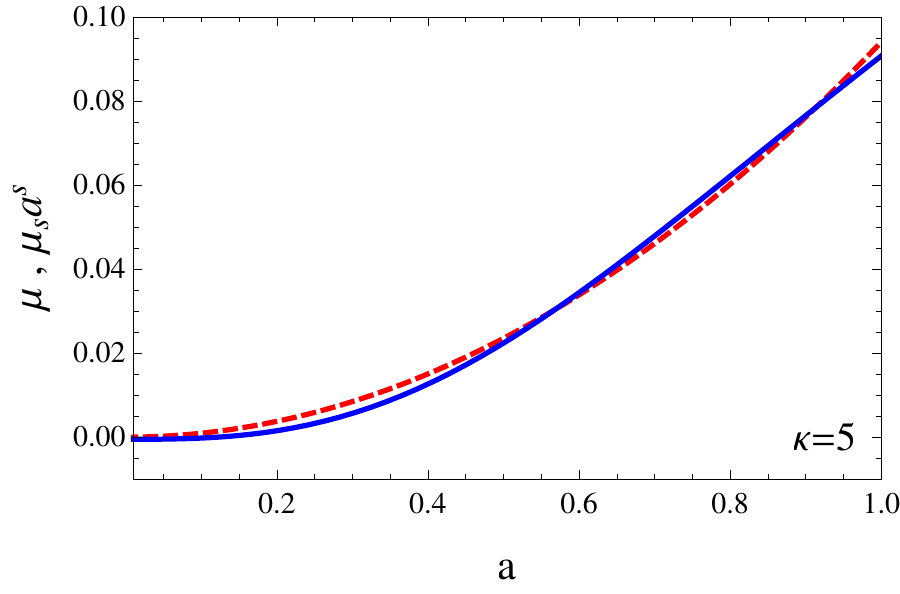}
\caption{\label{fig:N3N3b} Left panel: $\mu(z;k)$, as a function of the redshift $z$, for  
$\kappa = 0.1$ (red dashed) $\kappa=1$ (brown dot-dashed)
and $\kappa =5$ (blue solid line).  The curves for $\kappa = 1$ and  $\kappa=5$ are almost indistinguishable on this scale.
Right panel: the function $\mu$ for $\kappa=5$ (blue solid line), plotted against the scale factor $a$, and compared to the function $\mu(a)=\mu_sa^s$ with
$\mu_s=0.094 $ and $s=2$ (red dashed line). }
\end{figure}

\begin{figure}[t]
\centering
\includegraphics[width=0.45\columnwidth]{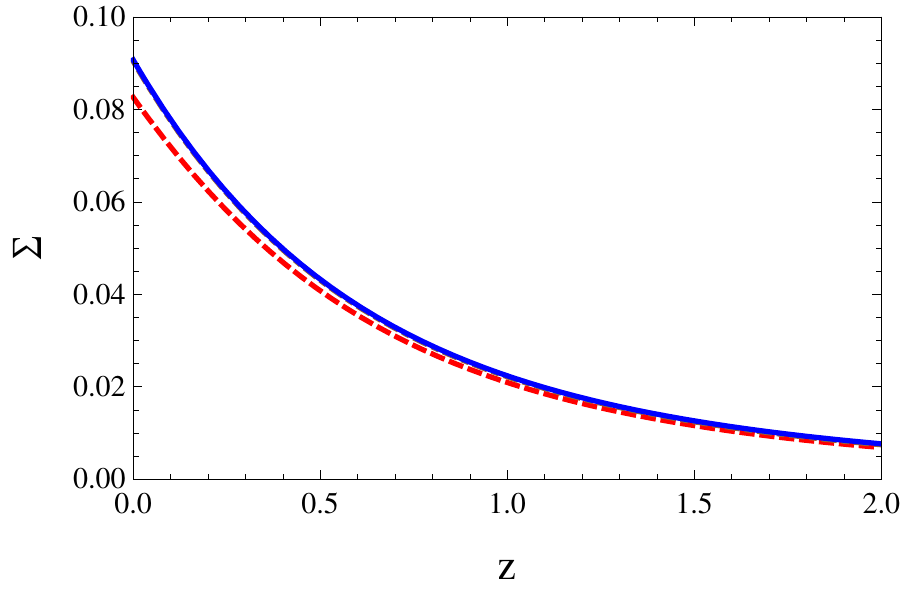}
\includegraphics[width=0.45\columnwidth]{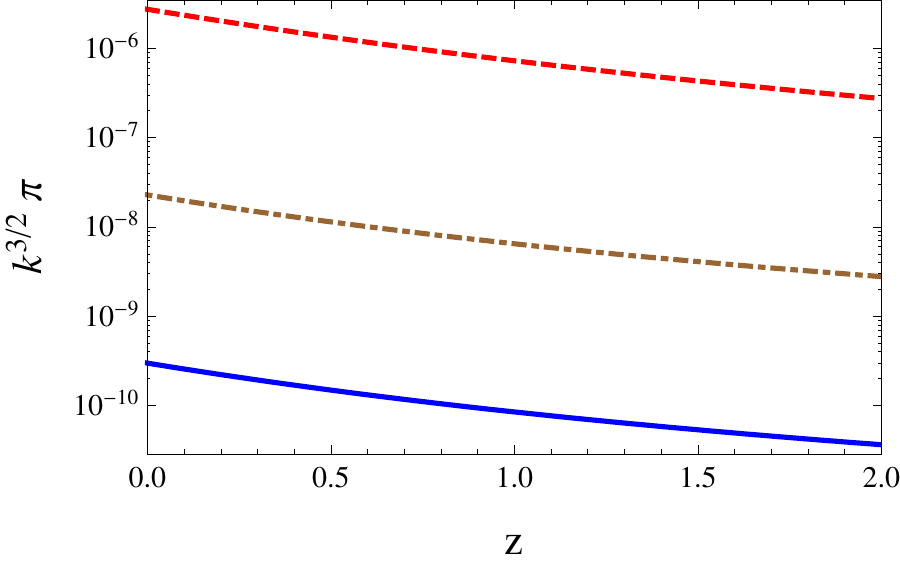}
\caption{\label{fig:N4N5}Left panel:   $\Sigma(z;k)$  as a function of the redshift $z$, for  
$\kappa = 0.1$ (red dashed) $\kappa=1$ (brown dot-dashed)
and $\kappa =5$ (blue solid line).
Right  panel: the same for $k^{3/2}\pi(z;k)$,}
\end{figure}

\begin{figure}[t]
\centering
\includegraphics[width=0.45\columnwidth]{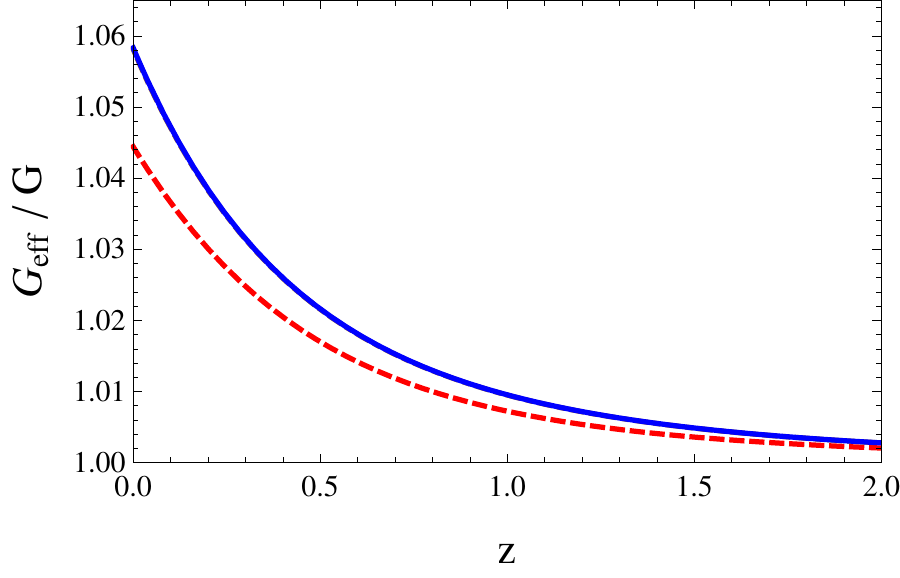}
\includegraphics[width=0.45\columnwidth]{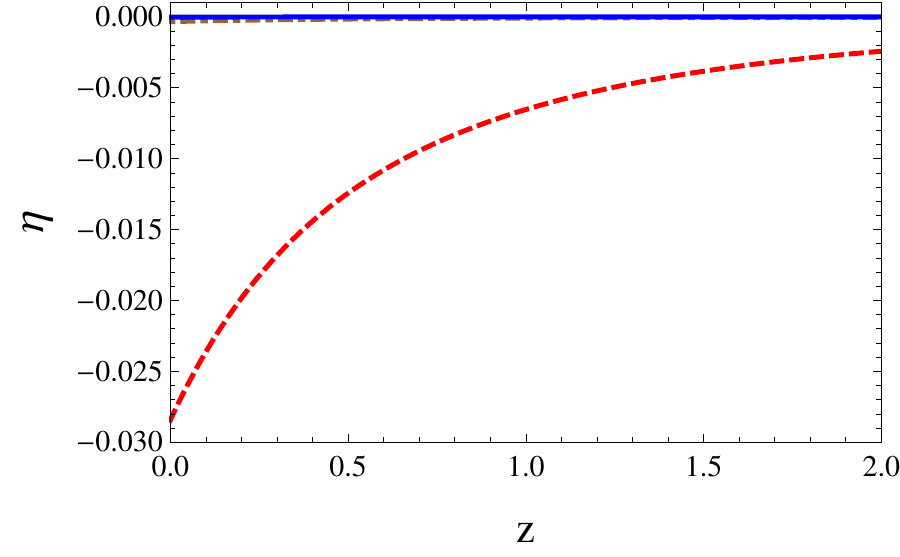}
\caption{\label{fig:N1N2} Left panel: $G_{\rm eff}(z;k)/G$, as a function of the redshift $z$, for  
$\kappa = 0.1$ (red dashed) $\kappa=1$ (brown dot-dashed)
and $\kappa =5$ (blue solid line).  The curves for $\kappa = 1$ and  $\kappa=5$ are almost indistinguishable on this scale.
Right panel: the same for $\eta(z;k)$.}
\end{figure}

We introduce $\kappa \equiv k/k_{\rm eq}$, where
$k_{\rm eq}=a_{\rm eq} H_{\rm eq}$ is the wavenumber of the mode that enters the horizon at matter-radiation equilibrium and, to illustrate our numerical results, we 
use as reference values $\kappa = 0.1$, $\kappa=1$ and $\kappa =5$. Since $k_{\rm eq}\simeq 0.014 \, h_0/{\rm Mpc}\simeq 42 H_0$,   $\kappa =5$ means $k/H_0\simeq 210$. This mode entered  inside the horizon already during RD and is   in  range given in \eq{rangekH0}. Thus, for this mode
we must recover the analytic results obtained in the  sub-horizon limit in Sect.~\ref{sect:analytic}.
The mode with $\kappa=1$ reentered at matter-radiation equality, and is still in the range
given in \eq{rangekH0}.
In contrast,  the mode with $\kappa=0.1$ (i.e. $k/H_0\simeq 4$)  was  outside the horizon during RD and most of MD, and re-entered at $z\simeq 1.5$. Overall, these three values of $k$ illustrate well  the $k$ dependence of the results. 

\subsection{Metric perturbations:  $\Psi$, $(\mu,\Sigma)$, $(G_{\rm eff},\eta)$}

The evolution of $\Psi(x;k)$ is shown in Fig.~\ref{fig:F1F2} for these three values of $\kappa$. Up to the present time, $x=0$, the evolution is quite similar to that in $\Lambda$CDM. Taking for instance the case $\kappa=1$, we see that this mode is constant (and negative) during RD, when it is outside the horizon, grows at the RD-MD transition, near $x_{\rm eq}\simeq -8.1$, and then becomes constant again, at a less negative value, during MD, where it is inside the horizon. We also see that it starts evolving again when DE starts to dominate, and in the far future it will go to zero.  

The ratio of $\Psi$ in our model to $\Psi_{\rm GR}$, which is defined as the value computed assuming GR and $\Lambda$CDM, defines $1+\mu(a;k)$, see \eq{defmu}. The
quantity $\mu(a;k)$ is shown on the left panel of
fig.~\ref{fig:N3N3b}, as a function of the redhsift $z$. We see that, for the sub-horizon modes relevant to structure formation, such as $\kappa=5$ or $\kappa =1$, $\mu(a;k)$ is basically independent of $k$, and $\mu(z=0.5)\simeq 0.04$. This is a particularly interesting value because, on the one hand, it is  well within the present observational limits
given in \eq{DPsiSimpson} and, on the other hand, is sufficiently large to be detectable in future surveys such as {\sc Euclid}.  From
the right-panel of  Fig.~\ref{fig:N3N3b} we see that  in the recent epoch  $\mu(a)$  is well reproduced by the parametrization used in ref.~\cite{Song:2010fg},
\be\label{muas}
\mu(a)=\mu_s a^s\, ,
\ee
with the values $\mu_s=0.094 $ and $s=2$.\footnote{Observe that the quantity that we call $1+\mu$ is called $\mu$ in~\cite{Song:2010fg}.
In any case, with both definitions the relation between $\Psi$ and $\Psi_{\rm GR}$, for sub-horizon modes, is 
$\Psi=[1+\mu_s a^s]\Psi_{\rm GR}$. Also, the exact value of $s$ that fits best the function $\mu (a)$ depends of course on the range $[a_{\rm min},a_{\rm max}]$ used to perform the fit.}
Future surveys such as {\sc Euclid} are expected to measure the parameter $\mu_s$ with great precision. In \cite{Song:2010fg} the forecast for {\sc Euclid} on the error $\sigma(\mu_s)$, for fixed cosmological parameters, is $\sigma(\mu_s)=0.0046$ for $s=1$ and 
$\sigma(\mu_s)=0.014$ for $s=3$. For our model we therefore  expect an accuracy of order $1\%$ or better on $\mu_s$, which would be largely sufficient to test our prediction
$\mu_s\simeq 0.09$.

The quantity $\Sigma(z;k)$, relevant for weak lensing, is shown in the left panel of Fig.~\ref{fig:N4N5}. Again, we see that at $z\simeq 0.5$ the corrections to GR are of order $4\%$, therefore consistent with present data, but potentially detectable.
The right panel of 
fig.~\ref{fig:N4N5} shows $\pi(z;k) \equiv \Phi(z;k) + \Psi(z;k)$, which is proportional to the anisotropic stress.

In the left panel of fig.~\ref{fig:N1N2} we plot $G_{\rm eff}(z;k)/G$, as a function of the redshift $z$, for three different values of the comoving momentum  $k$. From the left panel in Fig.~\ref{fig:N1N2}
we see that the numerical result for $G_{\rm eff}(z;k)/G$ for $\kappa=5$  (and also for $\kappa =1$) agrees indeed very well with the large-$\hat{k}$ limit shown in Fig.~\ref{fig:Geff}.
Quantitatively, in $z=0$ we get $G_{\rm eff}(z=0,\kappa=5)/G\simeq  1.0583$, in perfect agreement with \eq{Gefflargek}. 
The lower-$k$ modes have a  value of $G_{\rm eff}(z;k)/G$  closer to one. 
In the right panel of fig.~\ref{fig:N1N2} we plot $\eta(z;k)$, which confirms that, in the large-$\hat{k}$ limit, $\eta$ is extremely close to zero. For  $\kappa = 5$, we have  $|\eta |<2\times 10^{-5}$.
These results confirm that $(1-\eta) G_{\rm eff}/G$, which according to
\eqs{dmpp}{muetaG} is
the quantity relevant for the growth of $\d_M$, for sub-horizon modes deviates only by about 2 to 4\%  from the
$\Lambda$CDM values, in the range of redshifts relevant for comparison with the data. Thus, the deviations are consistent with existing data, but 
potentially detectable in the near future.

\subsection{Matter perturbations}

\begin{figure}[t]
\centering
\includegraphics[width=0.45\columnwidth]{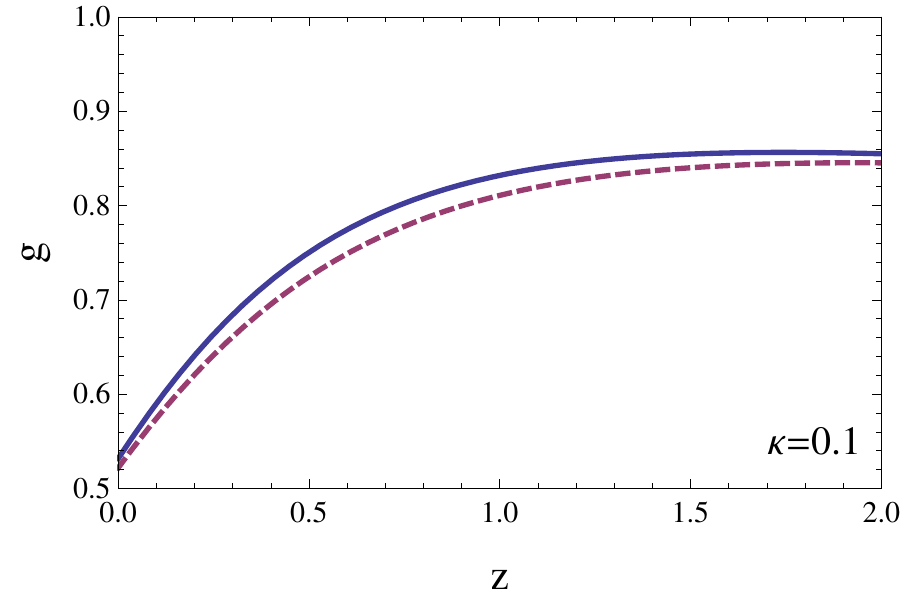}
\includegraphics[width=0.45\columnwidth]{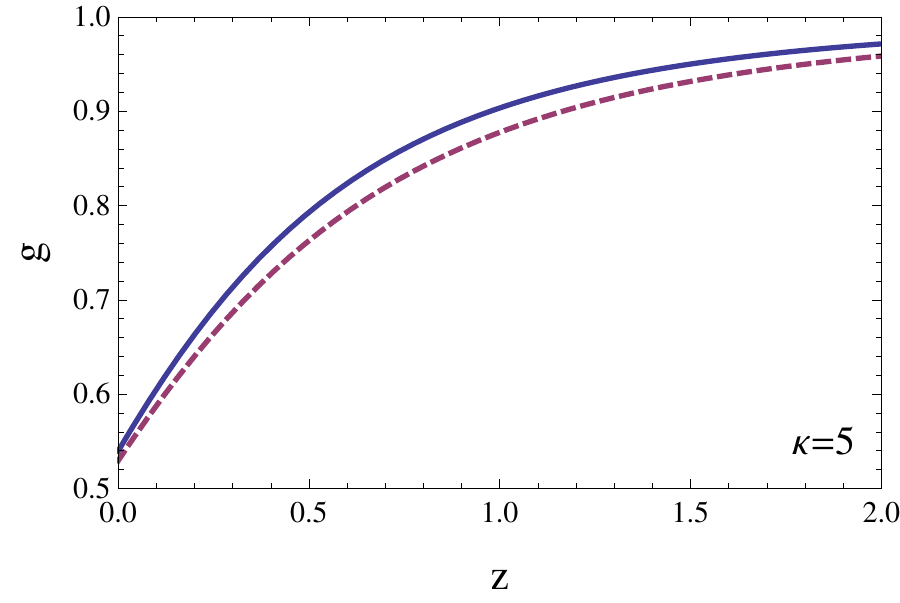}
\caption{\label{fig:G1G2} The   logarithmic growth rate in the $R\,\Box^{-2}R$ model
(blue solid line) compared to the same quantity in $\Lambda$CDM (purple dashed line).
Left panel: $\kappa=0.1$. Right panel: $\kappa=5$.}
\end{figure}

\begin{figure}[t]
\centering
\includegraphics[width=0.45\columnwidth]{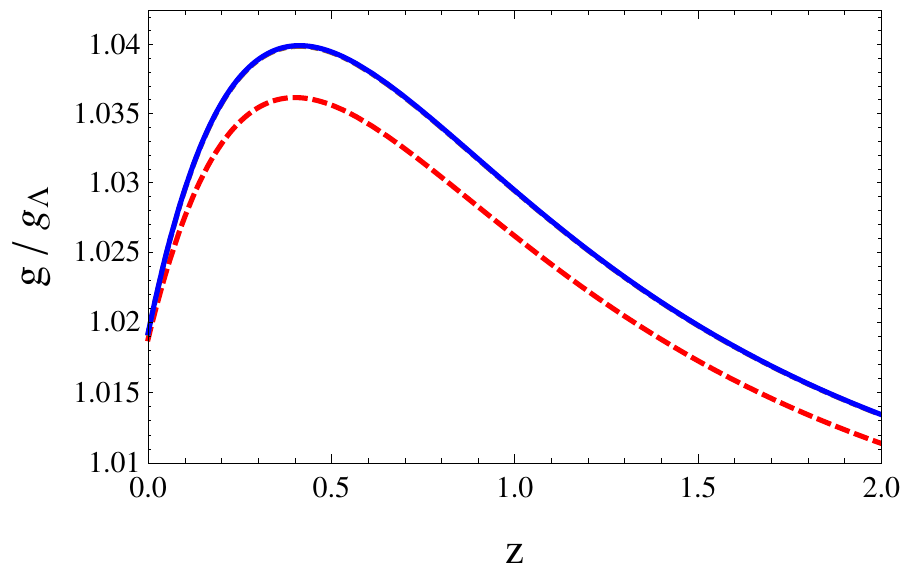}
\includegraphics[width=0.45\columnwidth]{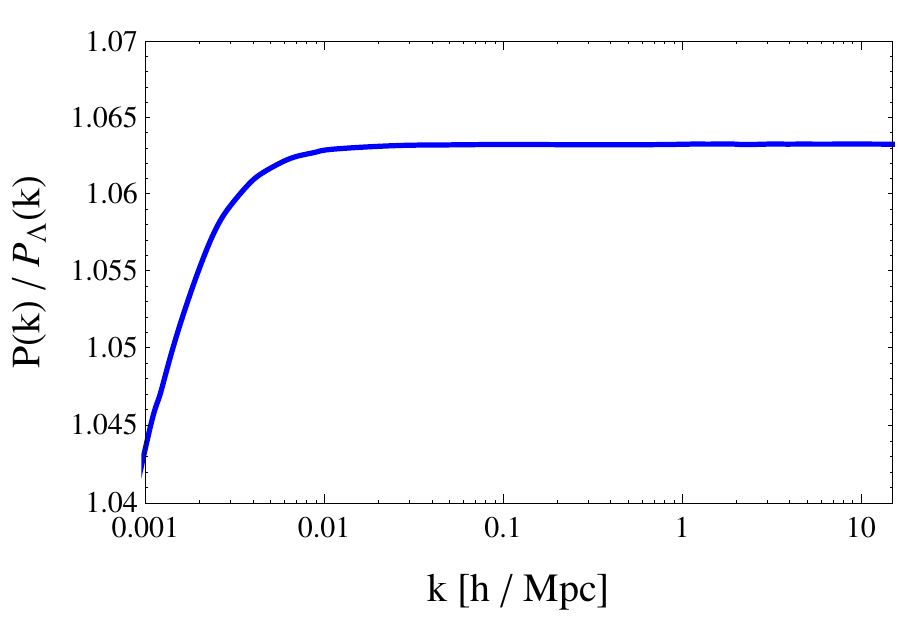}
\caption{\label{fig:G3} Left panel: the   ratio $g(z)/g_{\Lambda}(z)$
for $\kappa = 0.1$ (red dashed) $\kappa=1$ (brown dot-dashed)
and $\kappa =5$ (blue solid line).
Right panel: the  ratio of the linear matter power spectrum in the nonlocal model to 
the  linear matter power spectrum in 
$\Lambda$CDM, at $z=0$.}
\end{figure}

We next consider the matter perturbation $\d_M$. In fig.~\ref{fig:G1G2} we the plot  the logarithmic growth rate 
\be
g(z;k) \equiv \frac{d \log \delta_M}{d\ln a}
\ee
of the $R\,\Box^{-2}R$ model and we compare it with the same quantity  in
$\Lambda$CDM.  We show the results for 
$\kappa=0.1$ (left panel) and for $\kappa=5$ (right panel), while in the left panel of Fig.~\ref{fig:G3} we show the ratio $g(z)/g_{\Lambda}(z)$
for $\kappa=0.1$ and $\kappa=5$ (where $g_{\Lambda}$ is the quantity computed in GR with
$\Lambda$CDM). The ratio of the 
linear power spectrum of matter  in the nonlocal model, $P(k)$, to the linear power spectrum of matter  in 
$\Lambda$CDM, $P_{\Lambda}(k)$, at $z=0$, is shown in the right panel of Fig.~\ref{fig:G3}. We see that the two agree within a few percent, in the range of values of $k$ shown.\footnote{As already mentioned before, the comparison is performed taking the same values for the cosmological parameters, such as $\oma$, spectral index, etc. Of course, eventually in each model these parameters will be determined by a global fit to the data, and will not be the same for the two models.}

\begin{figure}[t]
\centering
\includegraphics[width=0.45\columnwidth]{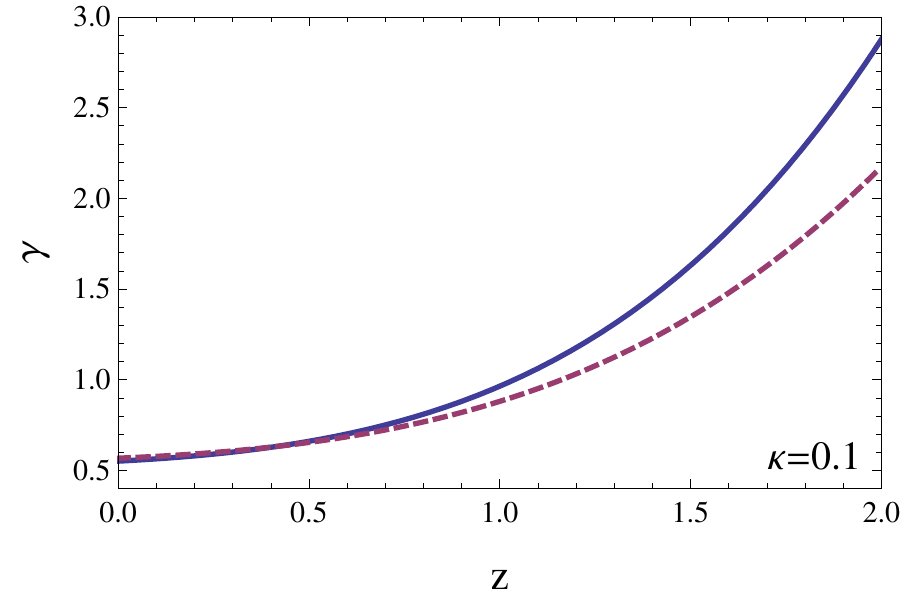}
\includegraphics[width=0.45\columnwidth]{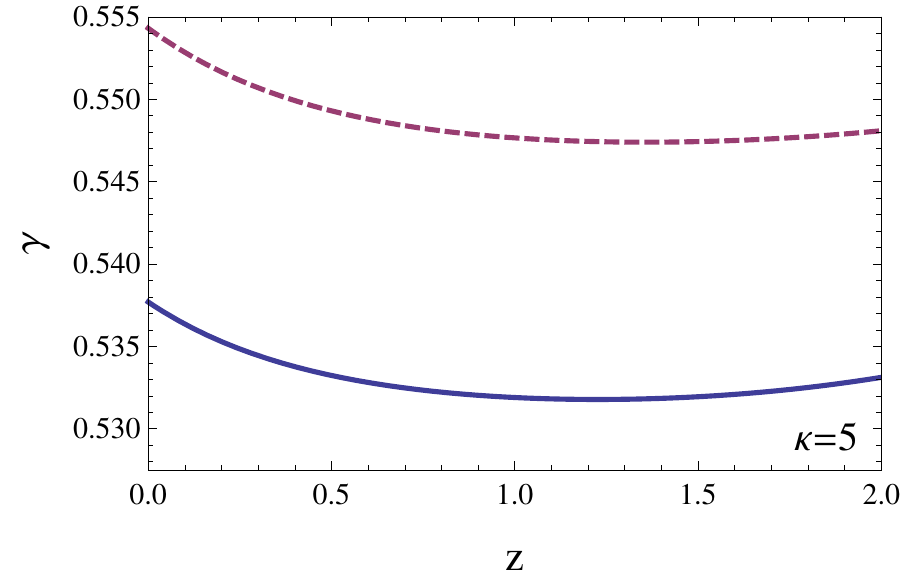}
\caption{\label{fig:G4G5} The    growth rate index $\gamma(z;k)$ in the $R\,\Box^{-2}R$ model
(blue solid line) compared to the same quantity in $\Lambda$CDM (purple dashed line).
Left panel: $\kappa=0.1$. Right panel: $\kappa=5$. These plots have been obtained setting $\oma= 0.3175$.}
\end{figure}

\begin{figure}[th]
\centering
\includegraphics[width=0.45\columnwidth]{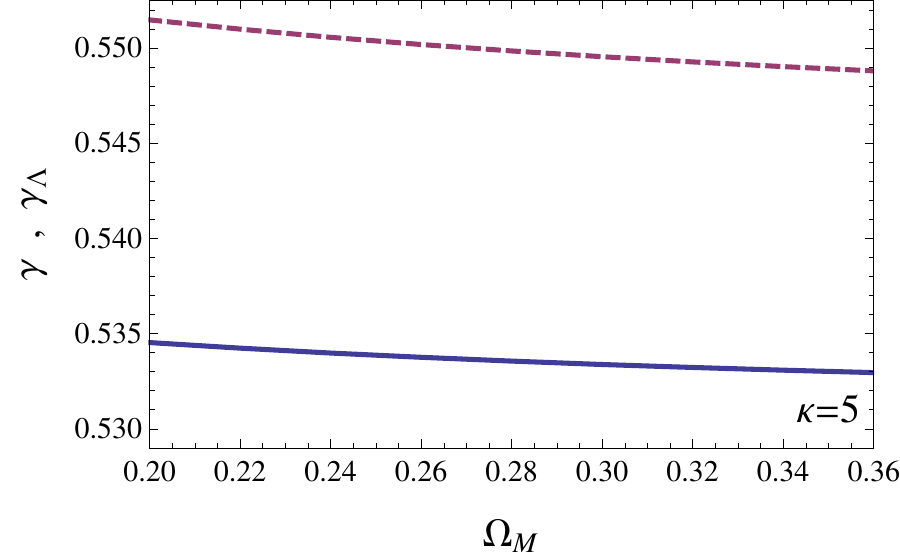}
\caption{\label{fig:GammavsOmegaM} The    growth rate index $\gamma(z;k)$ in the $R\,\Box^{-2}R$ model
(blue solid line) for $\kappa=5$ and $z=0.5$
as  a function of $\oma$, compared to the same quantity in $\Lambda$CDM (purple dashed line).}
\end{figure}

Another useful quantity is the growth rate index $\gamma(z;k)$ (not to be confused with the
parameter $\gamma =m^2/(9H_0^2)$), defined by
\be
g(z;k)=[\oma(z)]^{\gamma(z;k)}\, , 
\ee
where
\be
\Omega_M (z) =\frac{\rho_M(z)}{\rho_M(z) + \rho_R(z) + \rho_{DE}(z)}\, .
\ee
Observe that $\gamma(z;k)$ is also in principle  a function of the present value of the matter density
$\oma$.
We show it in Fig.~\ref{fig:G4G5} for $\oma= 0.3175$, again comparing with $\Lambda$CDM, for low and high momenta. 
In particular, for large $\hat{k}$, the result becomes independent of $\hat{k}$ and only weakly dependent on $z$, and we get
\be
\gamma\simeq 0.53\, ,
\ee
to be compared with the corresponding large-$\hat{k}$  value $\gamma\simeq 0.55$
in $\Lambda$CDM. As in $\Lambda$CDM, $\gamma$ remains almost constant
over the  range of redshift shown. Furthermore, just as in $\Lambda$CDM, the value of $\gamma(z)$ shows very little sensitivity to the value chosen for $\oma$ today. We illustrate this in Fig.~\ref{fig:GammavsOmegaM} where we show $\gamma$ for a fixed redshift $z=0.5$, as a function of $\oma$, both for the nonlocal model and for $\Lambda$CDM.

\clearpage

\subsection{Dark energy perturbations}

\begin{figure}[t]
\centering
\includegraphics[width=0.45\columnwidth]{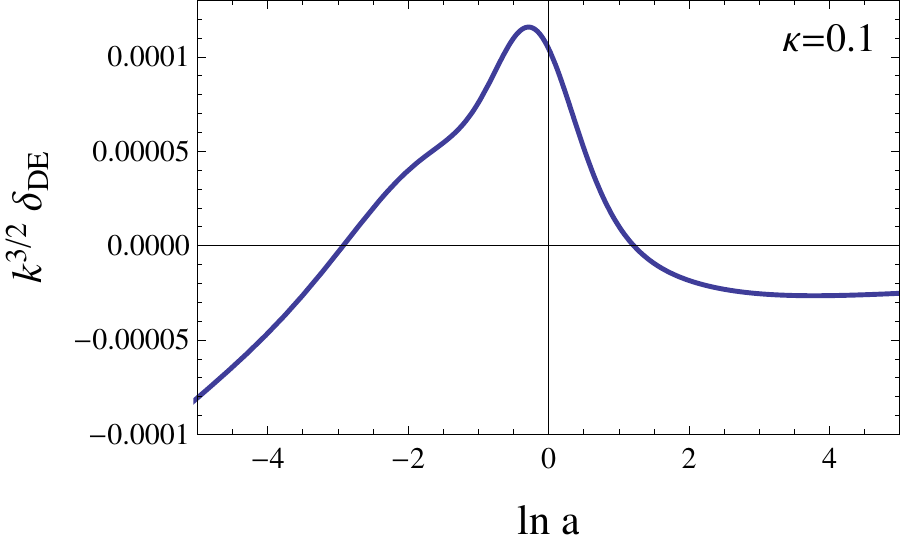}
\includegraphics[width=0.45\columnwidth]{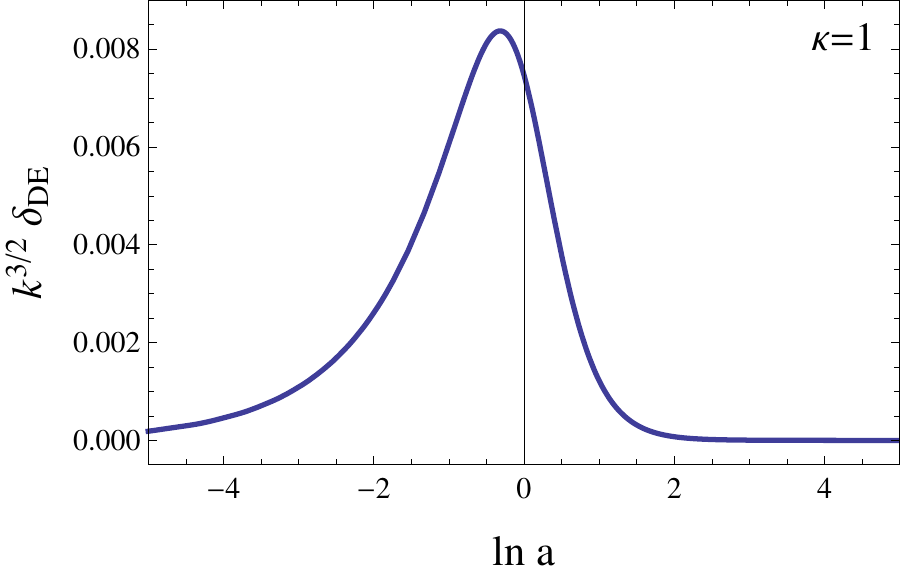}
\includegraphics[width=0.45\columnwidth]{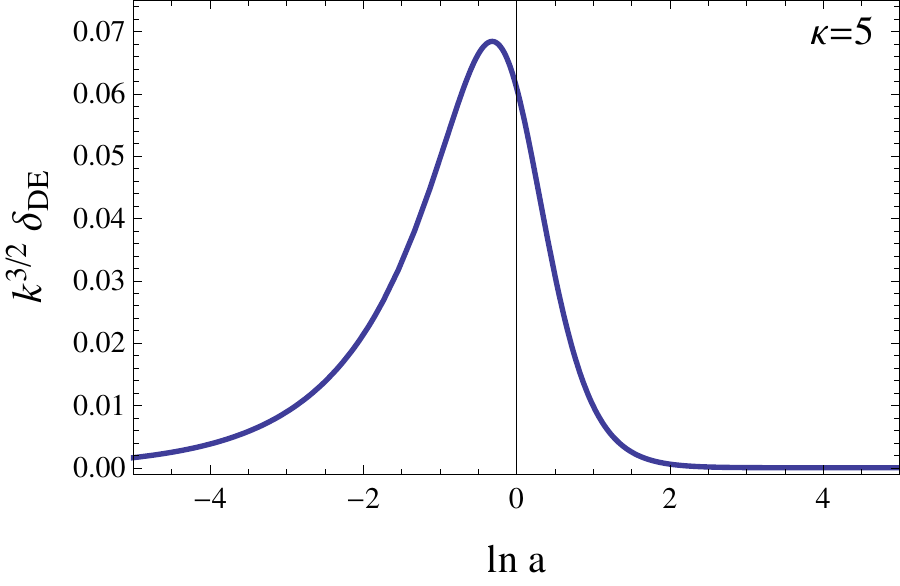}
\caption{\label{fig:DDE} The   quantity $k^{3/2}\d_{\rm DE}$ for $\kappa=0.1$ (top left),
$\kappa=1$ (top right) and $\kappa=5$ (bottom  panel).
}
\end{figure}

\begin{figure}[t]
\centering
\includegraphics[width=0.45\columnwidth]{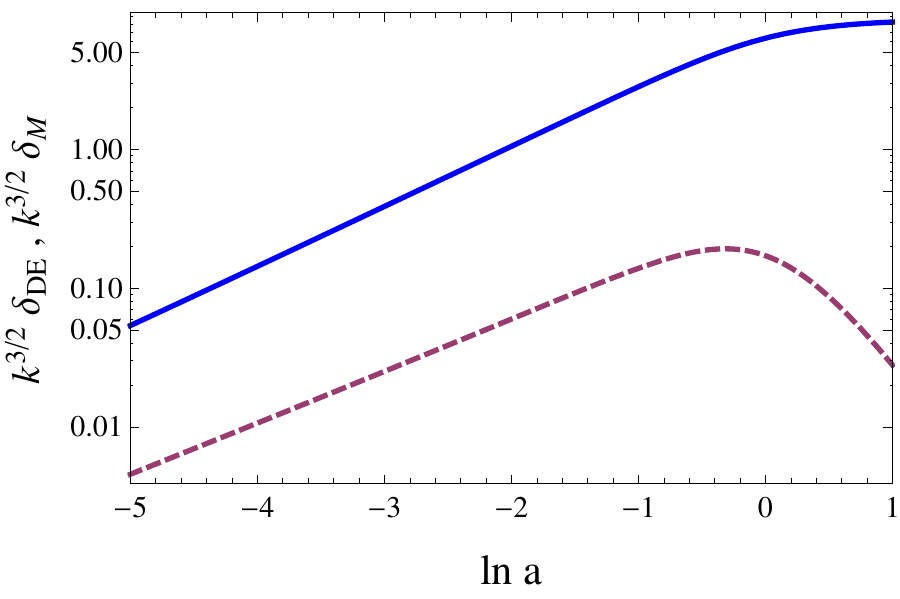}\includegraphics[width=0.45\columnwidth]{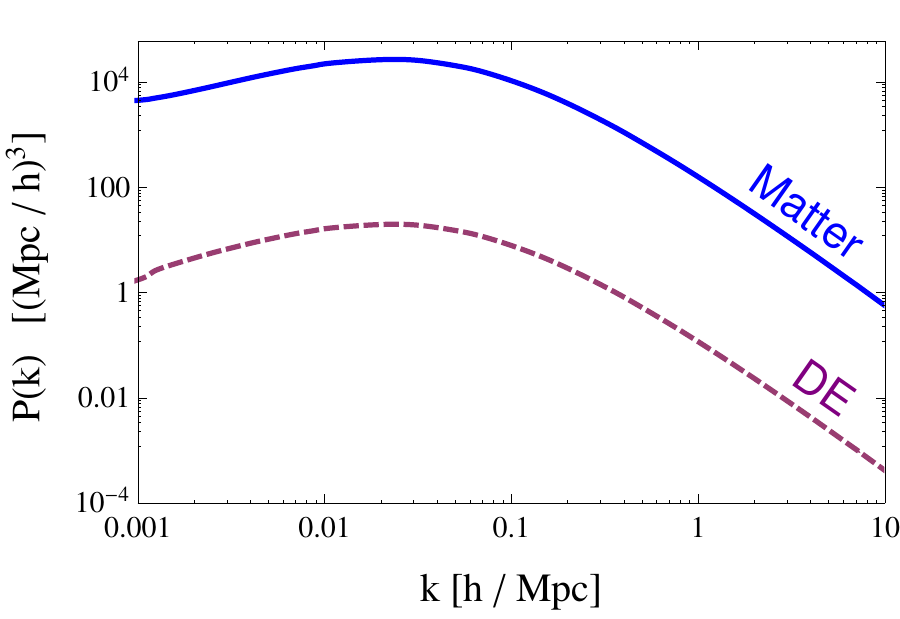}
\caption{\label{fig:PS} Left panel:  the matter perturbation $k^{3/2}\d_M$ (red, dashed)
 compared to the DE perturbation  $k^{3/2}\d_{\rm DE}$, for a mode with
$\kappa=20$, which becomes non-linear (of course, when $k^{3/2}\d_M$ becomes of order one, its evolution is no longer described by the linear theory). Right panel: the power spectrum of the linear matter and (effective) dark energy perturbations today. The dark energy perturbations
are much smaller than those of the dark matter on all scales relevant for cosmological structure formation.
}
\end{figure}

We next examine how the dark energy perturbation $\d_{\rm DE}$ evolves. 
In fig.~\ref{fig:DDE}  we show the evolution of $k^{3/2}\d_{\rm DE}(a;k)$
against $x=\ln a$, including the evolution in the far future, $x>0$, for the modes  with $\kappa=0.1$ (top left),
$\kappa=1$ (top right) and $\kappa=5$ (bottom panel).
We see that  in this model DE clusters, and $\d_{\rm DE}$  grows during the MD era. 
Increasing $\kappa$ also increases  the maximum value of $k^{3/2}\d_{\rm DE}(a;k)$ so, eventually, DE clustering becomes non-linear. However, this happens for values of $k$ where also the matter perturbations are non-linear. In 
the left panel of Fig.~\ref{fig:PS} we compare $\d_M$ and $\d_{\rm DE}$ for a mode with $\kappa=20$. We see that, for this mode, $k^{3/2}\d_M(a;k)$ becomes of order one around $x=-2$ (and then, of course, the subsequent evolution shown in the plot, and computed with linear theory, is no longer valid). When $k^{3/2}\d_M(a;k)$ becomes of order one, $k^{3/2}\d_{\rm DE}(a;k)$ is still in the linear regime, with a value of order 0.05. This means that  
structure formation in the nonlinear regime will still proceed in a first approximation as in $\Lambda$CDM, although with corrections due to the dark energy clustering.  
In the right panel of Fig.~\ref{fig:PS} we show the linear power spectrum of matter in the nonlocal model (blue solid line) and the linear power spectrum of dark energy perturbations (red dashed line), at redshift $z=0$.

 \subsection{Effective fluid description of dark energy perturbations}

\begin{figure}[t]
\centering
\includegraphics[width=0.45\columnwidth]{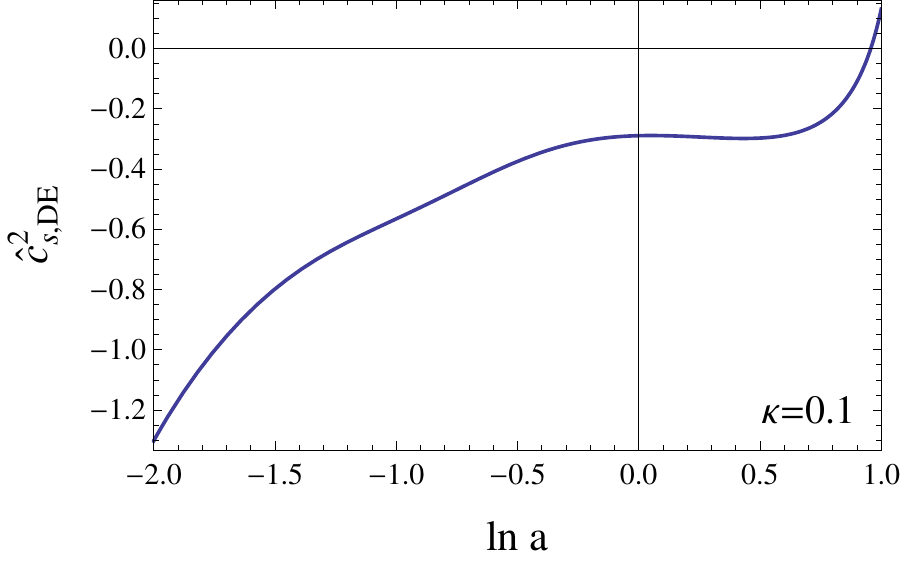}
\includegraphics[width=0.45\columnwidth]{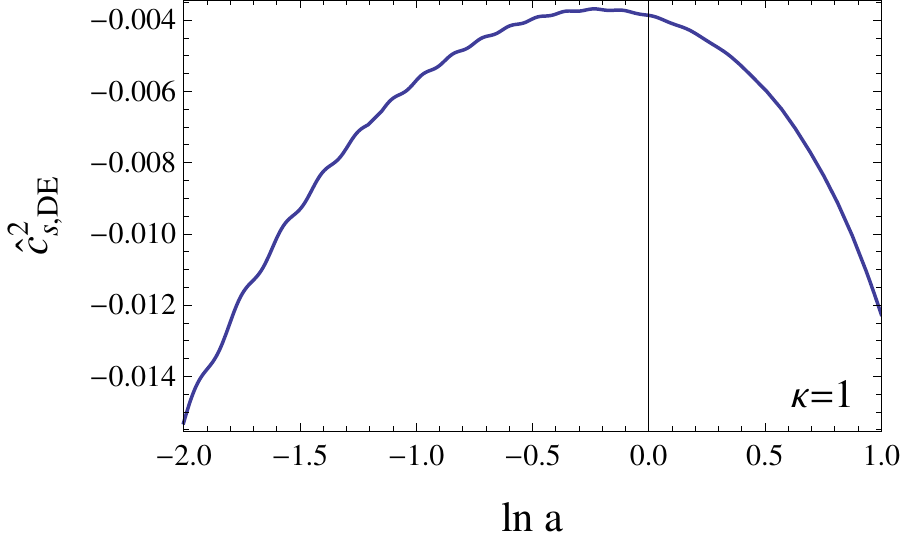}
\includegraphics[width=0.45\columnwidth]{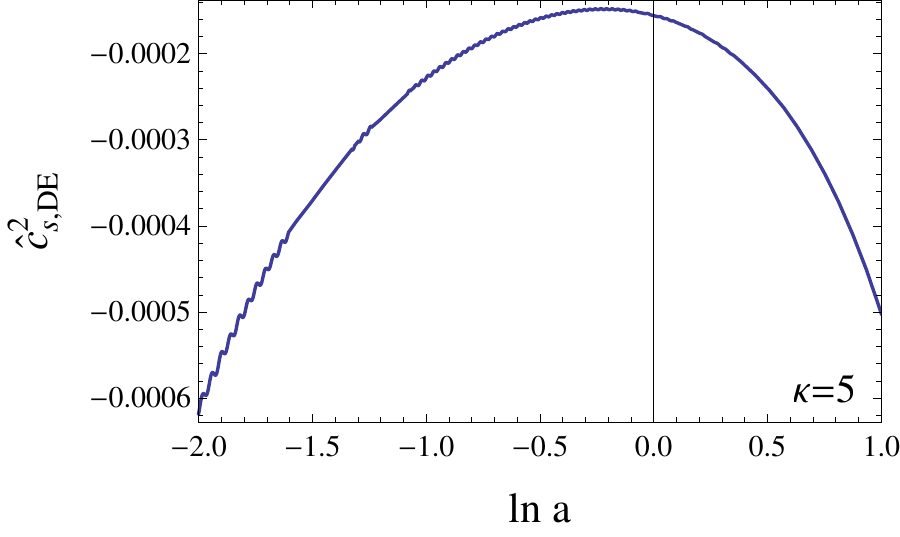}
\caption{\label{fig:CSDE} The DE  sound speed $\hat{c}^2_{s,\textsc{DE}}(x)$.}
\end{figure}

\begin{figure}[t]
\centering
\includegraphics[width=0.45\columnwidth]{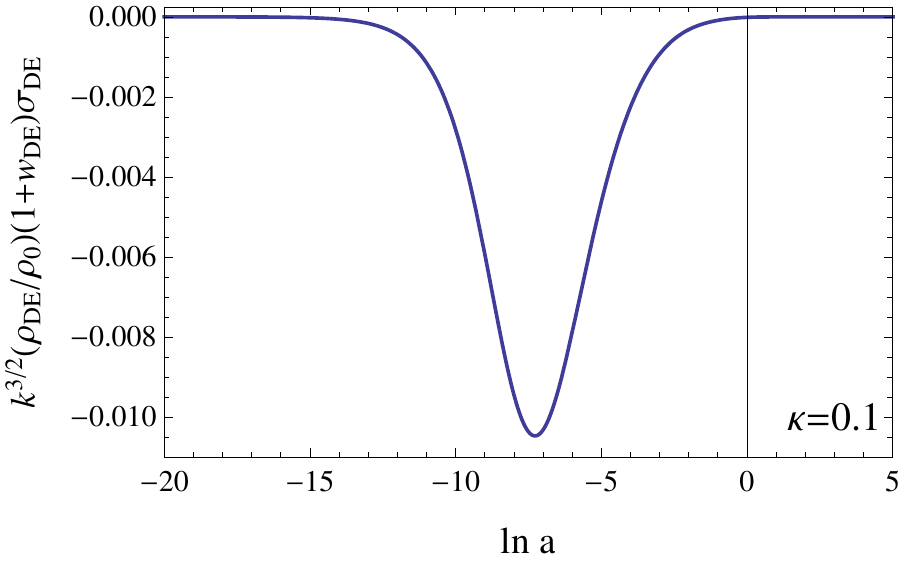}
\includegraphics[width=0.45\columnwidth]{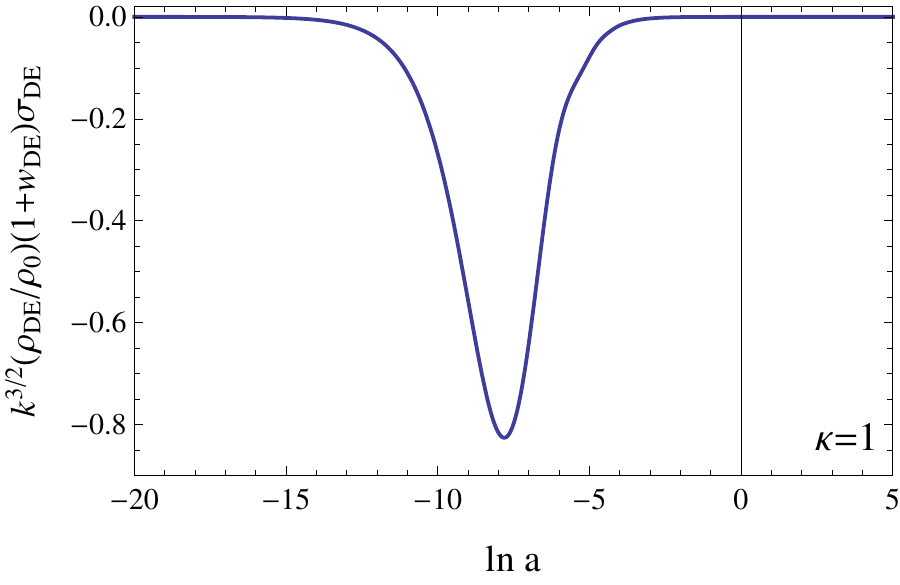}
\includegraphics[width=0.45\columnwidth]{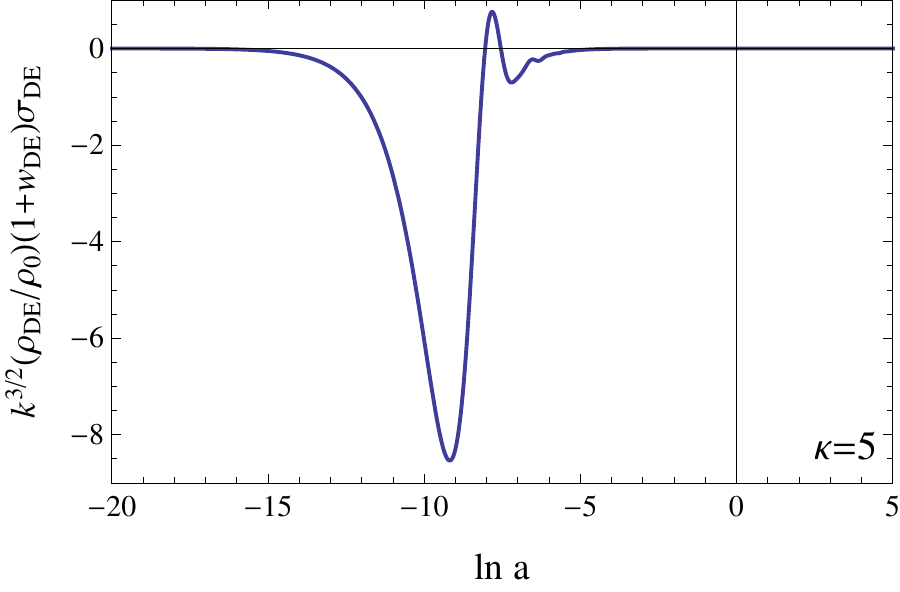}
\caption{\label{fig:SDE} The quantity 
$k^{3/2}(\rde/\rho_0)(1+w_{\rm DE})\sigma_{\textsc{DE}}(x)$, where $\sigma_{\textsc{DE}}(x)$
is the DE anisotropic stress.}
\end{figure}

\begin{figure}[t]
\centering
\includegraphics[width=0.45\columnwidth]{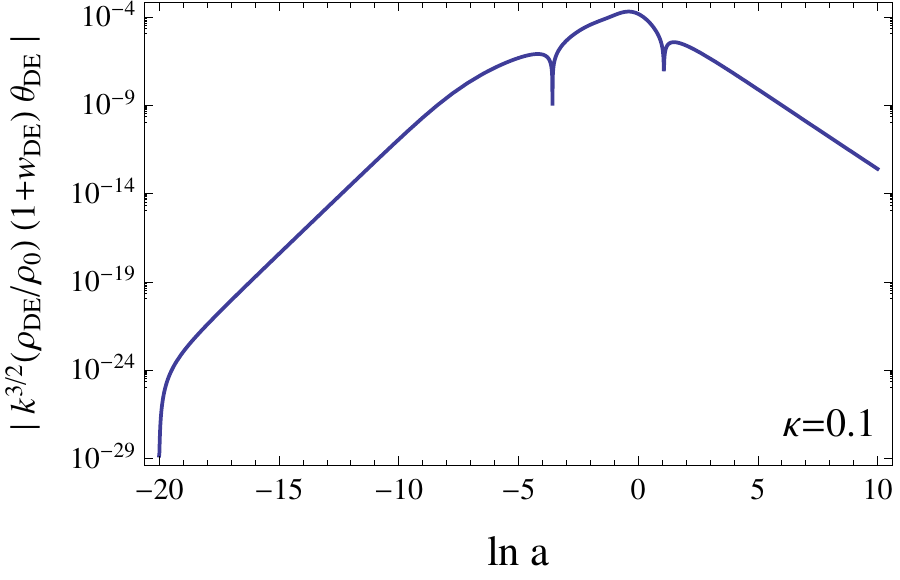}
\includegraphics[width=0.45\columnwidth]{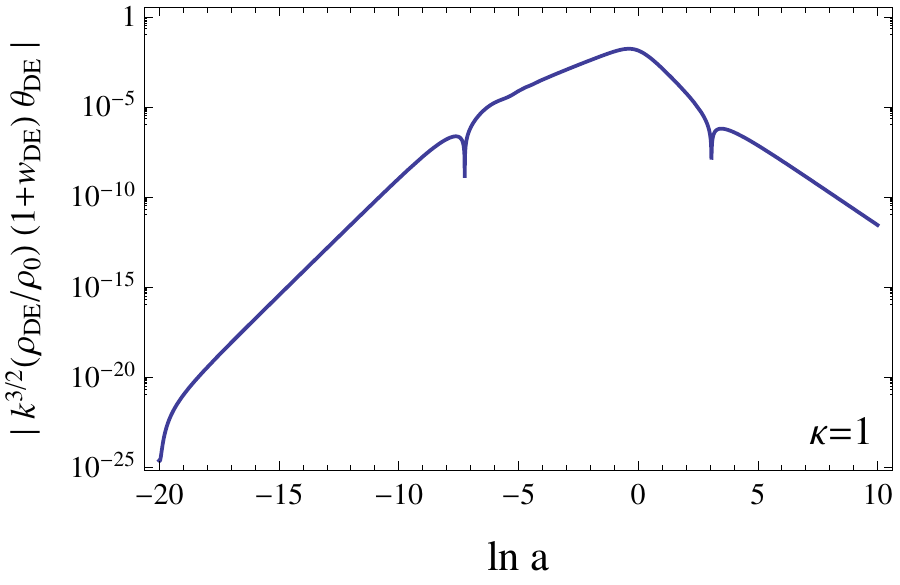}
\includegraphics[width=0.45\columnwidth]{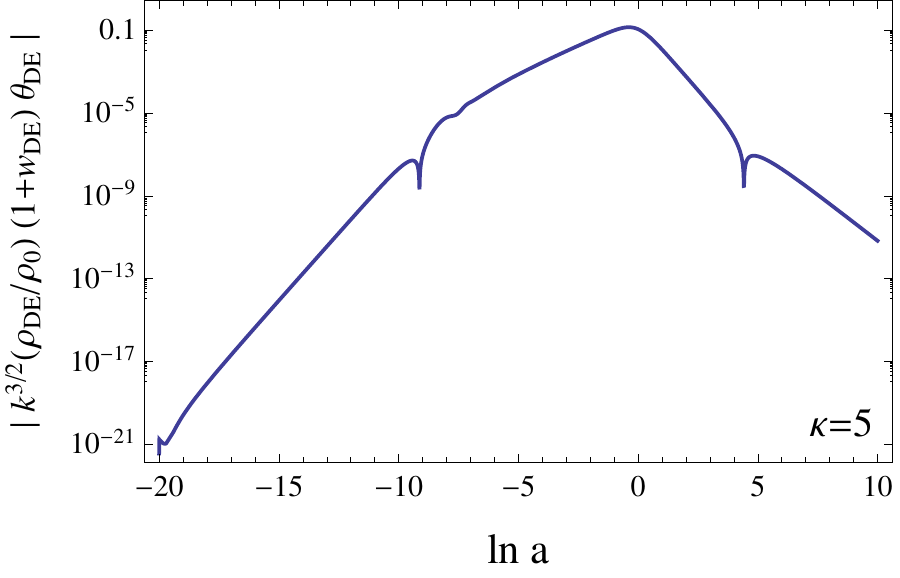}
\caption{\label{fig:TDE} The absolute value of  $k^{3/2}(\rde/\rho_0)(1+w_{\rm DE})\theta_{\textsc{DE}}(x)$.}
\end{figure}

Finally, in Figs.~\ref{fig:CSDE}-\ref{fig:TDE} we show for completeness the remaining quantities that characterize the DE perturbations as an effective fluid, i.e. the DE  sound speed $\hat{c}^2_{s,\textsc{DE}}(x)$,  the DE anisotropic stress $\sigma_{\textsc{DE}}(x)$ and the DE  velocity divergence $\theta_{\textsc{DE}}(x)$. We see in particular the both $\hat{c}^2_{s,\textsc{DE}}$, and 
$(1+w_{\rm DE})\sigma_{\textsc{DE}}$ are negative during MD. This helps us to understand the behavior of $\d_{\rm DE}$  shown in Fig.~\ref{fig:DDE}. Indeed, 
combining \eqs{conspert0}{consperti} for a generic conserved  DE fluid and taking the limit $\hat{k}\gg 1$ we get
\be\label{dddDE}
\d_{\rm DE}''+[2+\zeta-w_{\rm DE}+3(\hat{c}^2_{s,{\rm DE}}-w_{\rm DE})]\d_{\rm DE}'
+\hat{k}^2\[ \hat{c}^2_{s,{\rm DE}}\d_{\rm DE}+(1+w_{\rm DE})\sigma_{\rm DE}\]=0\, ,
\ee
(where we also made use of the fact that, because of Poisson equation,  $\Psi={\cal O}(1/\hat{k^2})$, and can be neglected with respect to $\sigma_{\rm DE}$ in the large-$\hat{k}$ limit). As we see from 
Figs.~\ref{fig:CSDE} and \ref{fig:SDE}, both 
$\hat{c}^2_{s,{\rm DE}}\d_{\rm DE}$ and $(1+w_{\rm DE})\sigma_{\rm DE}$ are negative (and comparable) and can therefore potentially drive an instability in $\d_{\rm DE}$. We see however from the plots of
$\d_{\rm DE}$ in Fig.~\ref{fig:DDE} that the instability is tamed when we enter in the epoch dominated by DE. Clearly, this is due to the accelerated expansion, that dilutes sufficiently fast the DE perturbations; in other words, in this regime the  friction term in \eq{dddDE}  wins over the terms $\hat{c}^2_{s,{\rm DE}}\d_{\rm DE}$
and $(1+w_{\rm DE})\sigma_{\rm DE}$.

\clearpage

\vspace{5mm}

\section{Conclusions}\label{sect:concl}

The introduction of nonlocal models such as those given in \eq{modelRT} and in \eq{S1} raises a number of interesting questions, both of conceptual nature, and on their viability as cosmological models.

At the  conceptual level, as  extensively discussed in \cite{Maggiore:2013mea,Foffa:2013sma,Kehagias:2014sda} and as we have mentioned in the Introduction, the crucial point is that such nonlocal equations of motion, involving a retarded propagator (which is necessary in order to ensure causality), cannot be taken as the equations of motion of a fundamental nonlocal quantum field theory. Rather, they must be understood as effective classical theories. In this direction, the main open problem is to understand if and how such nonlocal theories can be
obtained with some form of classical or quantum smoothing  from a more fundamental (and local) quantum  theory, see the discussion in the Introduction and in   \cite{Maggiore:2013mea}.

Another important set of questions, which was the focus of the present paper, concerns the phenomenological viability of such theories. As shown
in \cite{Kehagias:2014sda}, the nonlocal models  (\ref{modelRT}) and (\ref{S1}) 
recover all successes of GR at solar system and lab scales.  
In this paper we have worked out the cosmological perturbations of the nonlocal model
(\ref{S1}) (as well as of the model (\ref{modelRT}), see the Appendix). The main results that we have obtained can be summarized as follows.

\begin{itemize}

\item The cosmological perturbations are well-behaved.  Of course, cosmological perturbations are always unstable, even in $\Lambda$CDM. For instance, the dark matter perturbation $\d_M$ on sub-horizon scales in MD grows as $a$ and eventually become non-linear already in $\Lambda$CDM.  The issue is therefore whether the growth of the perturbation in the nonlocal model is sufficiently close to that of $\Lambda$CDM to be consistent with the observations, which is indeed the case in both the nonlocal models that we have studied.

\item A nonlocal model such as the $R\,\Box^{-2}R$ model defined  in \eq{S1}  is  remarkably predictive. In terms of a single parameter $m$ (that replaces the cosmological constant in $\Lambda$CDM), it predicts a whole set of functions of the redshift. At the background level, it gives a pure prediction for the dark energy equation of state parameter $w_{\rm DE}$ as a function of $z$. The result is shown 
in Fig.~\ref{fig:wDE}. Equivalently, it predicts the time  evolution  of the dark energy density, see Fig.~\ref{fig:rhoDE}. In particular, the EOS turns out to be phantom. With the usual parametrization (\ref{ChevLind}) near the recent epoch, we get $w_0 \simeq -1.14$ and  $w_a = 0.08$, with exact values depending on $\oma$, see Fig.~\ref{fig:wvsOM}, but still ranging over a relatively narrow set of values. Varying $\oma$ over the rather broad range $\oma\in [0.20,0.36]$,
$w_0$ remains within the relatively narrow interval $[-1.165,-1.135]$, while
$w_a\in [0.07,0.11]$. Similar considerations hold for the model defined in \eq{modelRT}, see the appendix.

\item At the perturbation level, the model fully predicts the energy density perturbations, pressure perturbations, anisotropic stress and velocity divergence, all as a function of redshift and of momentum, that fully characterize the DE perturbations as an effective fluid. From these, we can derive other quantities more readily comparable to the observation. In particular, structure formation is mostly affected by the function $\mu(a;k)$ defined by
\eq{defmu} while lensing is affected by the function $\Sigma(a;k)$ defined in \eq{defSigma}. We find that, for the modes relevant to observations, these function are to a very good approximation scale-invariant, i.e. independent of $k$. Our prediction for $\mu$ and $\Sigma$ as a function of redshift are given in
Figs.~\ref{fig:N3N3b} and \ref{fig:N4N5}. We find that the widely used parametrization $\mu(a)=\mu_s a^s$ fits  well our numerical results, and we predict  $\mu_s=0.094 $ and $s=2$. For the  
growth rate index $\gamma(z;k)$, we find again that, for the relevant modes, it is scale-independent, and the result is given in Fig.~\ref{fig:G4G5}. As in $\Lambda$CDM, it is in a first approximation independent also of $z$, and has the value $\gamma\simeq 0.53$, to be compared with 0.55 in $\Lambda$CDM.

\item Comparison with structure formation shows that the difference between this model and 
$\Lambda$CDM is small with respect to the present observational errors. The model therefore fits structure formation at a level that at present is statistically indistinguishable from $\Lambda$CDM. This is a non-trivial result. For instance, the non-local model proposed in \cite{Deser:2007jk} has been ruled out at the $8\sigma$ level by the comparison with structure formation~\cite{Dodelson:2013sma}. We have also verified that the nonlocal models fit the SNa~Ia data from the JLA set, again with an accuracy which is statistically indistinguishable from $\Lambda$CDM, see Table~\ref{tab:bgfit}. It is particularly interesting the fact that the deviations from $\Lambda$CDM are sufficiently small, so that the model passes these tests, but still sufficiently large to allow a clear distinction to be made with near-future surveys.

\end{itemize}

We believe that these nonlocal models can provide a new and interesting line of attack to the dark energy problem and, to the least, they can be a useful benchmark against which we can compare $\Lambda$CDM.

\vspace{5mm}
\noindent
{\bf Acknowledgments.}  We thank Michele Mancarella and Ermis Mitsou for useful discussions.
The work of YD, SF, MK and MM is supported by the Fonds National Suisse.
NK thanks  the D\'epartement de Physique Th\'eorique   of Geneva University for the hospitality during part of this work.

\appendix

\section{Cosmological perturbations in the $\gmn\iBox R$ model}\label{app:gmnR}

In this appendix we summarize the main results of a similar analysis performed for the
$\gmn\iBox R$ model, defined by 
\eq{modelRT} \cite{Maggiore:2013mea}. Some of our results overlap with those recently presented in
\cite{Nesseris:2014mea}. 
We define  again
$U= -\iBox R$, as in \eq{Udef}, and  for this model we also introduce
$\Smn=-U\gmn =\gmn\ \iBox R$. The extraction of the  transverse part can be performed  exploiting the fact that, in a generic Riemannian manifold, any symmetric tensor $\Smn$ can be decomposed as
\be\label{defStran}
S_{\mu\nu}=
S_{\mu\nu}^{\rm T}+\frac{1}{2}(\n_{\mu}S_{\nu}+\n_{\nu}S_{\mu})\, , 
\ee
with
$\n^{\mu}S_{\mu\nu}^{\rm T}=0$ \cite{Deser:1967zzb,York:1974}. 
In terms of $U$ and $S_{\mu}$, the original nonlocal equation (\ref{modelRT}) can be rewritten as
\bees
G_{\mu}^{\nu} + \frac{m^2}{3}  \[  U \delta_{\mu}^{\nu} + \frac{1}{2} 
\( \nabla_\mu S^\nu + \nabla^\nu S_\mu  \) \] &=& 8 \pi G T^{\mu}_{\nu}, \label{ModEE}\\
- \square_g U &=& R, \label{Loca} \\
 \nabla_\nu \( \nabla_\mu S^\nu + \nabla^\nu S_\mu  \) 
 &=& -2\partial_\mu U\label{DEEMcons}\, .
\ees
where \eq{DEEMcons} has been obtained by taking the divergence of \eq{defStran}. Observe that, since the left-hand side of 
\eq{modelRT} is transverse by construction, the energy-momentum  $T^{\mu}_{\nu}$ is automatically conserved. The background evolution of this model has been discussed in \cite{Maggiore:2013mea}. To study the cosmological perturbations in the scalar sector, we write again the metric as
in \eq{defPhiPsi} and we expand the auxiliary fields as
\be
U=\bar{U}+\d U\, ,\qquad S_{\mu}=\bar{S}_{\mu} +\d S_{\mu}\, .
\ee
In FRW the background value
$\bar{S}_{i}$ vanishes because there is no preferred spatial direction, but of course the perturbation $\d S_i$ is a dynamical variable. As with any vector, we can decompose it into a transverse and longitudinal part,
$\d S_i=\d S_i^{\rm T}+\pa_i (\d S)$
where $\pa_i(\d S_i^{\rm T})=0$. Since we  restrict here to scalar perturbations, we only retain $\d S$, and write $\d S_i=\pa_i (\d S)$. Thus (as already found in \cite{Kehagias:2014sda,Nesseris:2014mea}) in this model the scalar  perturbations  are given by $\Psi,\Phi,\d U,\d S_0$ and $\d S$, i.e. there is one more scalar variable compared to the $R\,\Box^{-2}R$ model.
We find convenient to trade $S_0$ and $S$ for the variables $V=H_0a^{-1}S_0$ and 
$Z=H_0^2 S$.
Linearizing the Einstein equations, going in momentum space, and using again a prime to denote the derivative with respect to $x=\ln a$, we get 
\bees
&&\hspace{-8mm}\hat{k}^2 \Phi + 3(  \Phi' - \Psi) = \frac{3}{2h^2\rho_0}
\[ \delta \rho + \gamma \rho_0 \( \delta U - h  \delta V' + 2 h \Psi \bar{V}' + h  \Psi' \bar{V} \) \], \label{EE1xRT}\\
&&\hspace{-8mm} \hat{k}^2 (\Phi' - \Psi ) = - \frac{3}{2h^2\rho_0}
\[ \bar{\rho} (1+w) \hat{\theta} + \hat{k}^2 \gamma \rho_0 
\(  h^2 \delta Z - \frac{h^2}{2} \delta Z' + h \Psi \bar{V} - \frac{h}{2} \delta V \) \], \label{EE2xRT} \\
&&\hspace{-8mm} \hat{k}^2 (\Psi + \Phi) =  \frac{9 }{2h^2\rho_0} 
\[ \bar{\rho} (1+w) e^{2x}\sigma +  \frac{2}{3} \hat{k}^2 \gamma\rho_0  h^2 \delta Z  \],  \label{EE3xRT}\\
&& \hspace{-8mm} \Phi'' + (3+\zeta)  \Phi' -  \Psi' - (3+2\zeta) \Psi + \frac{\hat{k}^2}{3} (\Phi + \Psi) \nonumber \\
&& = - \frac{3}{2h^2\rho_0}
\[ \delta p - \gamma \rho_0 \( \delta U - h (  \Phi' - 2 \Psi ) \bar{V} - h \delta V - \frac{\hat{k}^2}{3} h^2 \delta Z \)  \]\, . \label{EE4xRT}
\ees
The linearization of the equations for the auxiliary fields now gives 
\bees
&& \hspace{-10mm} \delta U'' + (3 + \zeta)  \delta U' + \hat{k}^2 \delta U = 2 \hat{k}^2 (\Psi + 2 \Phi)   + 6 ( \Phi'' +(4+\zeta)  \Phi' ) - 6 \[ \Psi' + 2(2+ \zeta) \Psi \] \nonumber \\
&& \hspace{5mm}
+ 2 \Psi  \bar{U}'' +
\[ 2 \Psi (3+\zeta) + (\Psi' - 3  \Phi') \]  \bar{U}'\,\label{AppU} \\
&&\hspace{-10mm} \delta V'' + (3+\zeta)  \delta V' 
+ \frac{\hat{k}^2}{2} h (  \delta Z' - 4 \delta Z )- h^{-1} \d U'  =  2 \Psi  \bar{V}''
 + \[ 2(3+\zeta) \Psi + 3( \Psi' -  \Phi') \]  \bar{V}' \nonumber \\
&&  \hspace{5mm}
 + \[ \Psi'' + (3 + \zeta)  \Psi' + 6 \Phi' \] \bar{V}  - \[ (1/2) \hat{k}^2 - 3 \] \( \delta V - 2 \Psi \bar{V} \)\, ,
\label{AppV} \\
&& \hspace{-10mm}
 \delta Z'' + (1+\zeta)\delta Z' + 2 \(  \hat{k}^2 - (3+\zeta) \) \delta Z =2h^{-2}\d U\nn\\
&&\hspace{5mm}-h^{-1}\[  \delta V' + 5 \delta V - 4 \Psi  \bar{V}' - 2 (  \Psi' -  \Phi' + 4 \Psi ) \bar{V} \]
\label{AppZ}\, .
\ees
The linearized Einstein  equations can again be recast in the form (\ref{EE1})-(\ref{EE4}), where now
\begin{align}
\delta \rho_{\rm DE}& \equiv \gamma \rho_0 \big( \delta U - h  \delta V' + 2 h \Psi  \bar{V}' + h\bar{V}  \Psi'  \big), \\
\bar{\rho}_{\rm DE} (1+w_{\rm DE}) \hat{\theta}_{\rm DE} &\equiv \hat{k}^2 \gamma \rho_0 \bigg(  h^2 \delta Z - \frac{h^2}{2} \delta Z' + h \Psi \bar{V} - \frac{h}{2} \delta V \bigg),\\
\bar{\rho}_{\rm DE} (1+w_{\rm DE}) \sigma_{\rm DE} &\equiv  \frac{2}{3} \hat{k}^2 \gamma\rho_0 e^{-2x} h^2 \delta Z, \\
\delta p_{\rm DE} &\equiv  - \gamma \rho_0 \bigg( \delta U - h (  \Phi' - 2 \Psi ) \bar{V} - h \delta V - \frac{\hat{k}^2}{3} h^2 \delta Z \bigg)\, .
\end{align}
Taking the large-$\hat{k}$ limit in \eqst{AppU}{AppZ}
we get $\d U=2\Psi+4\Phi$ while $\d V={\cal O}(1/\hat{k}^2)$ and
$\d Z={\cal O}(1/\hat{k}^2)$. Then,
from \eq{EE1xRT} we see that, for sub-horizon modes,
\be\label{GeffGRTlargek}
\frac{G_{\rm eff}}{G}=1+{\cal O}\(\frac{1}{\hat{k}^2}\)\, ,
\ee
in agreement with ref.~\cite{Nesseris:2014mea}. Comparing with \eq{Geff1suk2}, we see that in the
$\gmn\iBox R$ model the deviations in structure formation, with respect to $\Lambda$CDM, are even smaller than in the $R\,\Box^{-2}R$ model. The $\gmn\iBox R$ model is also closer to $\Lambda$CDM as far as the background evolution is concerned, since it predicts a value of $w_0\simeq -1.04$ (again, with a slight dependence on $\oma$), compared to $w_0\simeq -1.14$ for the $R\,\Box^{-2}R$ model. Since furthermore the dark energy perturbations are proportional to $(1+w_{\rm DE})$, and vanish as 
$w_{\rm DE}\ra -1$, in general we expect that the predictions  of the $\gmn\iBox R$ model will be intermediate between the prediction of $\Lambda$CDM and that of the $R\,\Box^{-2}R$ model.

\begin{figure}[t]
\centering
\includegraphics[width=0.45\columnwidth]{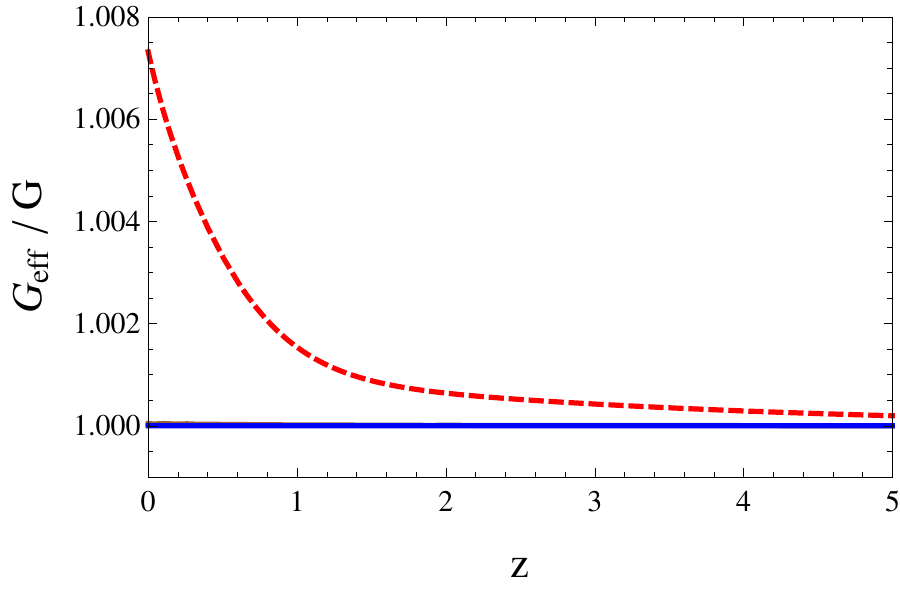}
\includegraphics[width=0.45\columnwidth]{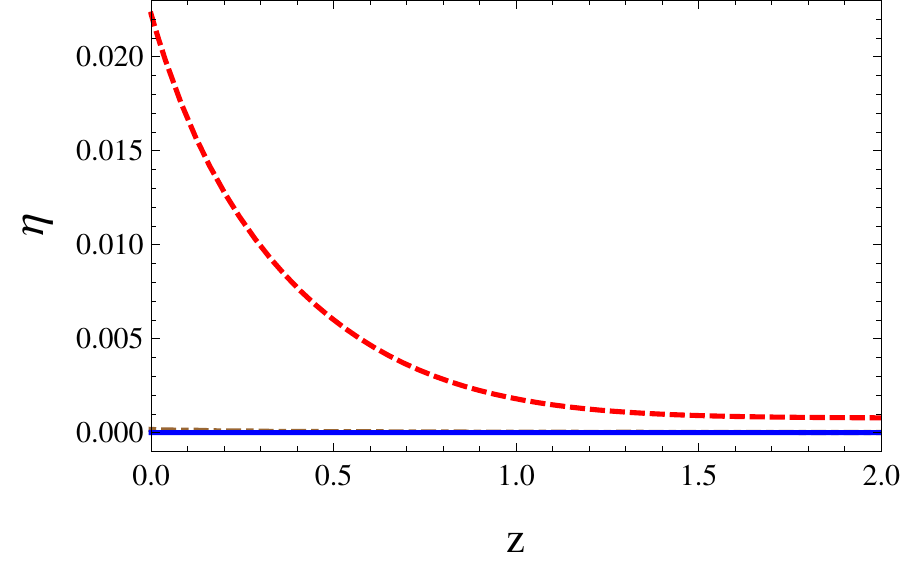}
\caption{\label{fig:N1N2app} Left panel: $G_{\rm eff}(z;k)/G$, as a function of the redshift $z$, for  
$\kappa = 0.1$ (red dashed) $\kappa=1$ (brown dot-dashed)
and $\kappa =5$ (blue solid line) for the $\gmn\iBox R$ model. 
The curves for $\kappa=1$ 
and $\kappa =5$ are indistinguishable on this scale.
Right panel: the same for $\eta(z;k)$.}
\end{figure}

In Fig.~\ref{fig:N1N2app} we show $G_{\rm eff}(z;k)/G$, and $\eta$ from the full numerical integration in the $\gmn\iBox R$ model. As expected from \eq{GeffGRTlargek}, for large $\hat{k}$  (i.e. $\kappa\,\gsim \, 1$) 
$G_{\rm eff}(z;k)/G$ is equal to one with great accuracy, and even for smaller values, such as 
$\kappa=0.1$, $G_{\rm eff}(z;k)$ is equal to  $G$ to better than $1\%$.
This should be compared with fig.~\ref{fig:N1N2}  and with \eq{Geff1suk2} for the
$R\,\Box^{-2}R$ model, where instead for large $k$ we have $G_{\rm eff}(z=0;k)/G\simeq 1.06$. We see again that the $\gmn\iBox R$ model is closer to $\Lambda$CDM, with respect to  the
$R\,\Box^{-2}R$ model. We also see that, in the two models, the sign of $\eta$ differs for low $k$, while $\eta$ is totally negligible in both cases for large $k$.

\begin{figure}[t]
\centering
\includegraphics[width=0.45\columnwidth]{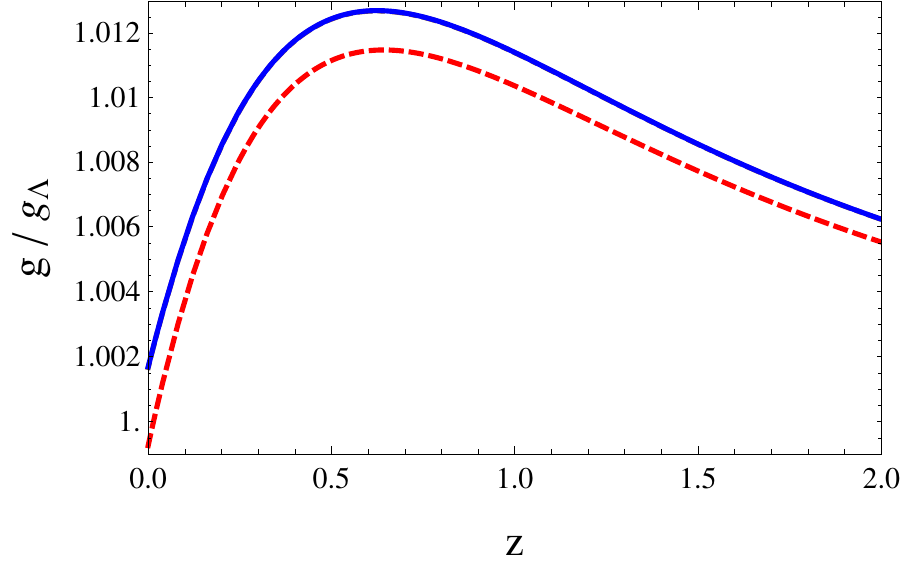}
\includegraphics[width=0.45\columnwidth]{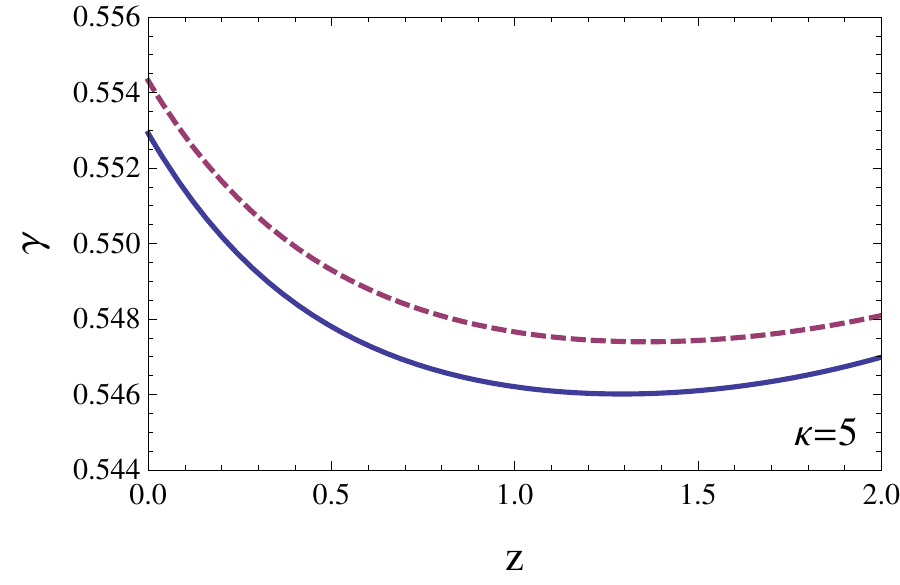}
\caption{\label{fig:G3app}
Left panel: the ratio of the logarithmic growth rates in the $\gmn\iBox R$ model
and  in $\Lambda$CDM,
for $\kappa = 0.1$ (red dashed) 
and $\kappa =5$ (blue solid line). Right panel: the    growth rate index $\gamma(z;k)$ in the $R\,\Box^{-2}R$ model
(blue solid line) compared to the same quantity in $\Lambda$CDM (purple dashed line), for $\kappa=5$.}
\end{figure}

In the left panel of Fig.~\ref{fig:G3app} we plot the ratio of the logarithmic growth rates in the $\gmn\iBox R$ model
and  in $\Lambda$CDM, to be compared with Fig.~\ref{fig:G3} for the $R\,\Box^{-2}R$ model,
while in the right panel we show the growth index. For the $\gmn\iBox R$ model, we find that
$\gamma$ is the same as in $\Lambda$CDM at the level of the first two digits. These results confirm that the perturbations in this model are quite close to that in $\Lambda$CDM, and the main difference is at the level of the background evolution, due to $w_0\simeq -1.04$. This also justifies the treatment of
\cite{Nesseris:2014mea}, where the model has been fitted to the CMB data using the shift parameter (which in principle assumes that the fluctuations are the same as in $\Lambda$CDM) rather than a full Boltzmann code. Such a treatment would however be less accurate for the $R\,\Box^{-2}R$ model.

\clearpage

\bibliographystyle{utphys}
\bibliography{myrefs_massive}

\end{document}

%% file: mydefs.tex
\newcommand{\nn}{\nonumber}

\newcommand{\scDE}{{\textsc{DE}}}

\newcommand{\iBox}{\Box^{-1}}

\renewcommand\({\left(}
\renewcommand\){\right)}
\renewcommand\[{\left[}
\renewcommand\]{\right]}

\newcommand\n{{\mbox {\boldmath $\nabla$}}}
\newcommand{\ra}{\rightarrow}

\def\lsim{\raise 0.4ex\hbox{$<$}\kern -0.8em\lower 0.62
ex\hbox{$\sim$}}

\def\gsim{\raise 0.4ex\hbox{$>$}\kern -0.7em\lower 0.62
ex\hbox{$\sim$}}

\def\lbar{{\hbox{$\lambda$}\kern -0.7em\raise 0.6ex
\hbox{$-$}}}

\newcommand\eq[1]{eq.~(\ref{#1})}
\newcommand\eqs[2]{eqs.~(\ref{#1}) and (\ref{#2})}

\newcommand\eqst[2]{eqs.~(\ref{#1})--(\ref{#2})}

\newcommand\pa{\partial}
\newcommand\p{\partial}

\newcommand\ee{\end{equation}}
\newcommand\be{\begin{equation}}
\def\bea{\begin{array}}
\def\eea{\end{array}}\def\ea{\end{array}}
\newcommand\ees{\end{eqnarray}}
\newcommand\bees{\begin{eqnarray}}
\def\nn{\nonumber}

\def\s{\sigma}
\def\g{\gamma}

\def\d{\delta}

\def\dslash{\hspace{-1mm}\not{\hbox{\kern-2pt $\partial$}}}
\def\Dslash{\not{\hbox{\kern-4pt $D$}}}
\def\pslash{\not{\hbox{\kern-2.1pt $p$}}}
\def\kslash{\not{\hbox{\kern-2.3pt $k$}}}
\def\qslash{\not{\hbox{\kern-2.3pt $q$}}}

\newcommand{\vk}{{\bf k}}
\newcommand{\vx}{{\bf x}}

\def\p1{{\bf p}_1}
\def\p2{{\bf p}_2}
\def\k1{{\bf k}_1}
\def\k2{{\bf k}_2}

\newcommand{\emn}{\eta_{\mu\nu}}

\newcommand{\gmn}{g_{\mu\nu}}

\newcommand{\hmn}{h_{\mu\nu}}

\newcommand{\bhmn}{\bar{h}_{\mu\nu}}

\newcommand{\Gmn}{G_{\mu\nu}}

\newcommand{\Tmn}{T_{\mu\nu}}
\newcommand{\Smn}{S_{\mu\nu}}

\newcommand{\dddM}{\kern 0.2em \raise 1.9ex\hbox{$...$}\kern -1.0em \hbox{$M$}}
\newcommand{\dddQ}{\kern 0.2em \raise 1.9ex\hbox{$...$}\kern -1.0em \hbox{$Q$}}
\newcommand{\dddI}{\kern 0.2em \raise 1.9ex\hbox{$...$}\kern -1.0em\hbox{$I$}}
\newcommand{\dddJ}{\kern 0.2em \raise 1.9ex\hbox{$...$}\kern-1.0em
\hbox{$J$}}
\newcommand{\dddcalJ}{\kern 0.2em \raise 1.9ex\hbox{$...$}\kern-1.0em
\hbox{${\cal J}$}}

\newcommand{\dddO}{\kern 0.2em \raise 1.9ex\hbox{$...$}\kern -1.0em
\hbox{${\cal O}$}}
\def\dddz{\raise 1.5ex\hbox{$...$}\kern -0.8em \hbox{$z$}}
\def\dddd{\raise 1.8ex\hbox{$...$}\kern -0.8em \hbox{$d$}}
\def\dddbd{\raise 1.8ex\hbox{$...$}\kern -0.8em \hbox{${\bf d}$}}
\def\ddbd{\raise 1.8ex\hbox{$..$}\kern -0.8em \hbox{${\bf d}$}}
\def\dddx{\raise 1.6ex\hbox{$...$}\kern -0.8em \hbox{$x$}}

\newcommand{\Sch}{Schwarzschild }

\newcommand{\ode}{\Omega_{\rm DE}}
\newcommand{\oma}{\Omega_{M}}
\newcommand{\ora}{\Omega_{R}}

\newcommand{\ola}{\Omega_{\Lambda}}

\newcommand{\rde}{\rho_{\rm DE}}